\documentclass[pdflatex,sn-mathphys-num,referee]{sn-jnl}

\usepackage{graphicx}%
\usepackage{multirow}%
\usepackage{amsmath,amssymb,amsfonts}%
\usepackage{amsthm}%
\usepackage{mathrsfs}%
\usepackage[title]{appendix}%
\usepackage{xcolor}%
\usepackage{textcomp}%
\usepackage{manyfoot}%
\usepackage{booktabs}%
\usepackage{algorithm}%
\usepackage{algorithmicx}%
\usepackage{algpseudocode}%
\usepackage{listings}%
\usepackage{anyfontsize}
\usepackage{enumitem}
\usepackage[per-mode=symbol]{siunitx}
\usepackage{subcaption}
\usepackage{threeparttable}
\usepackage[stable]{footmisc}
\usepackage{mathtools}

\newcommand{\bm}{{\mathbf{m}}}

\newcommand{\bA}{{\mathbf{A}}}

\newcommand{\bw}{{\mathbf{w}}}

\newcommand{\bh}{{\mathbf{h}}}

\newcommand{\revOne}[1]{\textcolor{black}{#1}}
\newcommand{\revTwo}[1]{\textcolor{black}{#1}}
\newcommand{\secRound}[1]{\textcolor{black}{#1}}

\begin{document}

\title[The trajectoRIR database]{The trajectoRIR Database: Room Acoustic Recordings Along a Trajectory of Moving Microphones}


\author*[1]{\fnm{Stefano} \sur{Damiano}}\equalcont{These authors contributed equally to this work.}\email{stefano.damiano@esat.kuleuven.be}

\author*[1]{\fnm{Kathleen} \sur{MacWilliam}}\email{kathleen.macwilliam@esat.kuleuven.be}\equalcont{These authors contributed equally to this work.}

\author[1]{\fnm{Valerio} \sur{Lorenzoni}}

\author[2]{\fnm{Thomas} \sur{Dietzen}}

\author[1]{\fnm{Toon} \sur{van Waterschoot}}


\affil[1]{\orgdiv{Dept. of Electrical Engineering (ESAT-STADIUS)}, \orgname{KU Leuven}, \city{Leuven}, \country{Belgium}}
\affil[2]{\orgdiv{Dept. of Electrical Engineering (ESAT-PSI), EAVISE}, \orgname{KU Leuven}, \city{ Sint-Katelijne-Waver}, \country{Belgium}}

\abstract{
Data availability is essential in the development of acoustic signal processing algorithms, especially when it comes to data-driven approaches that demand large and diverse training datasets. For this reason, an increasing number of databases have been published in recent years, including either room impulse responses (RIRs) or audio recordings during motion. In this paper we introduce the trajectoRIR database, an extensive, multi-array collection of both dynamic and stationary acoustic recordings along a controlled trajectory in a room. Specifically, the database contains moving-microphone recordings and stationary RIRs that spatially sample the room acoustics along an L-shaped trajectory. 
This combination makes trajectoRIR unique and applicable to a wide range of tasks, including sound source localization and tracking, spatially dynamic sound field reconstruction, auralization, and system identification.
The recording room has a reverberation time of $\SI{0.5}{\second}$, and the three different microphone configurations employed include a dummy head, with additional reference microphones located next to the ears, 3 first-order Ambisonics microphones, two circular arrays of 16 and 4 channels, and a 12-channel linear array. The motion of the microphones was achieved using a robotic cart traversing a $\SI{4.62}{\meter}$-long rail at three speeds: [0.2, 0.4, 0.8] $\si{\meter\per\second}$.
Audio signals were reproduced using two stationary loudspeakers. The collected database features 8648 stationary RIRs, as well as perfect sweeps, speech, music, and stationary noise recorded during motion. Python functions are provided to access the recorded audio and retrieve the associated geometric information.
}

\keywords{Room acoustic database, room impulse responses, moving microphone arrays, sound field reconstruction, dynamic acoustic scenes}

\maketitle

\section{Introduction}\label{sec:introduction}
Multi-microphone signal processing is becoming essential for an increasing number of room acoustics applications, ranging from telepresence, virtual room navigation, sound-zoning and spatial audio reproduction, to assistive hearing for hearing impaired people and robot audition. These applications inherently involve dynamic acoustic scenes in which both listeners (microphones) and emitters (sound sources) are free to move in the environment, usually being a room. In this context, several challenges arise, including room parameter estimation~\cite{macwilliamStatespaceEstimationSpatially2024}, sound source localization and tracking~\cite{diaz-guerraRobustSoundSource2021}, auralization~\cite{aliSource-timeDominantModeling2025},
speech enhancement~\cite{quanMultichannelLongTermStreaming2024}, echo cancellation~\cite{nophut2024velocity-controlled} and sound field \revTwo{estimation~\cite{katzbergCompressedSensingFramework2018}}. 
\revTwo{Broad data availability is a key requirement for developing and evaluating algorithms}, especially given the recent surge in machine and deep learning techniques that achieve state-of-the-art results when tackling the aforementioned challenges~\cite{vanWaterschootDeepDataDriven2025,diaz-guerraRobustSoundSource2021,quanMultichannelLongTermStreaming2024,zhengBATLearningReason2024,heDeepNeuralRoom2024}. 
\revTwo{Room acoustics research involves both stationary scenes, where sources and microphones are in fixed positions within a room, or dynamic scenes, where they are free to move in the space.} 

\revTwo{To target applications involving stationary settings, static audio recordings or room impulse responses (RIRs) are required, which can be either real or simulated. High-quality datasets containing music~\cite{nielsenSingleMultichannelAudio2014}, speech~\cite{eatonACEChallenge20142015,nielsenSingleMultichannelAudio2014,woodsRealworldRecordingDatabase2015,sheelvantRSL2019RealisticSpeech2019}, and babble or cocktail-party noise~\cite{woodsRealworldRecordingDatabase2015,dietzenMYRiAD2023} have been collected, but they fail to represent dynamic acoustic scenes.
Additionally, several recorded RIR datasets involving various microphone array configurations exist~\cite{stewartDatabaseOmnidirectionalBformat2010,nielsenSingleMultichannelAudio2014,hadadMultichannelAudioDatabase2014,eatonACEChallenge20142015,woodsRealworldRecordingDatabase2015,sheelvantRSL2019RealisticSpeech2019,cmejlaMIRaGeMultichannelDatabase2021,venkatakrishnanTampereUniversityRotated2021,koyamaMESHRIRDatasetRoom2021,mckenzie2021datasetspatialroomimpulse,zhaoRoomImpulseResponse2022,dietzenMYRiAD2023,miotelloHOMULARIRRoomImpulse2024,kujawskiMIRACLEMicrophoneArray2024, treybigHighSpatialResolution2024} and are widely used to generate synthetic data.} 
Simulations are, in fact, an inviting tool for generating large amounts of data under arbitrary acoustic conditions, including variations in array configurations, room parameters, signal types, and source or microphone trajectories. 
\revTwo{In room acoustics research, an established way to generate synthetic data is to convolve source signals with RIRs. This enables the creation of complex acoustic scenes involving a spatial distribution of sound sources. 
In this case, the adopted RIRs are either recorded, resulting in a higher realism at the cost of a more limited control over room parameters, or simulated, allowing to create arbitrary rooms and acoustic scenes at the price of a more limited physical accuracy.
} 
As a drawback, simulated environments often deviate from real-world conditions, which may result in algorithms that do not robustly generalize to real scenarios. 

\revTwo{To address use-cases involving dynamic settings, different datasets that capture dynamic acoustic scenes also exist, tailored for either audio~\cite{heDeepNeuralNetworks2018a,straussDREGONDatasetMethods2018,eversLOCATAChallengeAcoustic2020,politisDatasetReverberantSpatial2020a,politisSTARSS22DatasetSpatial2022,brunnstromExperimental2025} or multi-modal applications~\cite{lathoudAV163AudioVisualCorpus2005,deleforgeLatentlyConstrainedMixture2012,alameda-pinedaRAVELAnnotatedCorpus2013}. These corpora are designed for the task of sound source localization and tracking and include either speech, noise, or environmental sound events collected using different array geometries. These geometries comprise planar~\cite{eversLOCATAChallengeAcoustic2020,politisDatasetReverberantSpatial2020a,politisSTARSS22DatasetSpatial2022,brunnstromExperimental2025}, circular~\cite{lathoudAV163AudioVisualCorpus2005,brunnstromExperimental2025}, cubic~\cite{straussDREGONDatasetMethods2018}, and spherical~\cite{eversLOCATAChallengeAcoustic2020} arrays, as well as ambisonic microphones~\cite{politisDatasetReverberantSpatial2020a,politisSTARSS22DatasetSpatial2022}, dummy heads (DH)~\cite{deleforgeLatentlyConstrainedMixture2012,alameda-pinedaRAVELAnnotatedCorpus2013,eversLOCATAChallengeAcoustic2020}, and robot heads~\cite{heDeepNeuralNetworks2018a}.
\revOne{Since these datasets are designed for tasks involving motion, they do not contain stationary RIRs along the motion path. Such RIRs are essential for applications like sound field interpolation or room parameter estimation involving dynamic scenes, where synthetic data remains a common alternative~\cite{katzbergCompressedSensingFramework2018,macwilliamStatespaceEstimationSpatially2024}.} However, while there exist established procedures to fairly accurately synthesize these types of signals in static conditions under mild physical assumptions~\cite{allenImageMethodEfficiently1979,vorlanderAuralizationFundamentalsAcoustics2020}, synthesizing audio under motion is a challenging task as it requires either a precise spatiotemporal room acoustic model, or RIR interpolation from real recorded data.}

\revTwo{For several applications, including time-variant RIR estimation~\cite{macwilliamStatespaceEstimationSpatially2024}, spatially dynamic sound field reconstruction~\cite{brunnstromExperimental2025}, auralization~\cite{aliSource-timeDominantModeling2025}, and the evaluation of dynamic audio simulations, there is a clear need for recordings of audio under motion with corresponding stationary RIRs along the motion path. While some existing datasets provide either stationary or dynamic audio data, a database containing matching recordings for both remains unavailable.}

\revTwo{In an effort to bridge this gap and fuse the potential of RIR databases and audio recordings during motion, in this paper we introduce \emph{trajectoRIR}: an extensive, multi-array database of stationary and dynamic acoustic recordings performed along a trajectory in a reverberant room.} 
At the core of the trajectoRIR database lies a smoothed-L-shaped trajectory built using a rail system, on which a robotic cart is used to move microphones in a precise and reproducible manner. Both stationary RIRs, recorded at finely-spaced positions along the trajectory, and audio recordings during motion of the cart are included in the database. \revTwo{The recordings are obtained using three microphone array configurations: the first one (MC1) consists of a DH with in-ear omnidirectional microphones, two omnidirectional microphones located next to the ear canals, a 16-channel uniform circular array located around the DH at the same height of the ear canals, and a second 4-channel uniform circular array located above the head. The second one (MC2) is similar to MC1, only without the DH. The third configuration (MC3) consists of three first-order ambisonics (FOA) microphones and a 12-channel uniform linear array (ULA). These configurations have been chosen because they include standard microphone arrays that are widely employed in room acoustics and have been used for the collection of several existing RIR databases. This diversity thus enables the potential use of the proposed database in combination with other data collections, which makes it attractive for training data-driven algorithms~\cite{vanWaterschootDeepDataDriven2025}.}

\revTwo{The recordings were performed in the Alamire Interactive Laboratory (AIL)~\cite{lirias3940173} located in Heverlee, Belgium, at the Park Abbey where the recording room has a reverberation time of $\SI{0.5}{\second}$~\cite{dietzenMYRiAD2023}, and two static loudspeaker positions were used for the recordings.}
\revTwo{For stationary recordings, a total of 8648 RIRs were collected at equally spaced positions along the trajectory, with inter-position distances depending on the microphone configuration used. For the moving microphone recordings, the robotic cart was moved in a single direction at three different constant speeds in the walking speed range. During this movement, several signals were played from the speakers, including a piano piece, a drum beat, female speech, a white noise signal, and two perfect sweep signals covering different frequency ranges.}
A total of 108 multi-channel recordings were obtained, 36 per microphone array configuration. In addition, the ego-noise of the cart and rail system was recorded and added to the database in order to allow for the estimation of noise statistics and to promote the development of ego-noise reduction algorithms.
\revTwo{Alongside the recorded audio, extensive metadata is provided for all the recordings in accompanying csv files, including geometrical information, speed information for the moving recordings, and temperature data.} 

\revOne{To showcase the collected data, a systematic evaluation on the use-case of time-variant RIR estimation is introduced in Sec.~\ref{sec:use_cases}. In this evaluation, RIRs along the motion trajectory are estimated using (i) the (sparse) stationary RIR recordings collected at fixed positions; (ii) a moving-microphone recording; (iii) a combination of both. Results indicate that the RIRs reconstructed from a combination of stationary and moving audio data agree with recorded data the best overall, confirming the importance of the proposed database featuring matching recordings.}

The trajectoRIR database contains a total of $\SI{3.4}{\hour}$ of audio recorded at $\SI{48}{\kilo\hertz}$ in $\SI{24}{\bit}$, for a total size of $\SI{7.47}{\giga\byte}$. All material, together with source signals and Python scripts to use the data and retrieve geometrical information, is available at~\cite{trajectoRIR2024}.

The rest of the paper is organized as follows. In Sec.~\ref{sec:room_description}, we give an overview the AIL room. In Sec.~\ref{sec:recording_equipment}, we describe the recording equipment used. In Sec.~\ref{sec:spatial_setup}, we introduce the rail system and the robotic cart used to obtain the measurements. In Sec.~\ref{sec:mic_configuration}, we describe the microphone arrays and loudspeaker configurations used to capture the data. In Sec.~\ref{sec:recorded_signals}, we detail the input signals used for the recordings and provide information on the recording configurations for the stationary and moving scenarios, and in Sec.~\ref{sec:using_the_database} we give instructions to access the data, retrieve geometry information using the Python scripts, and provide examples of recorded signals. In Sec.~\ref{sec:use_cases} we present a systematic evaluation of the database on the use-case of time-variant RIR estimation. We finally summarize the database in Sec.~\ref{sec:conclusion}.

\section{Room description}\label{sec:room_description}
The AIL~\cite{lirias3940173} room, also used in the MYRiAD database~\cite{dietzenMYRiAD2023} and shown in Fig.~\ref{fig:room_setup_photo}, is a laboratory located in the Saint Norbert's gate of the Park Abbey in Heverlee, Belgium. The laboratory is approximately a shoebox-shaped room of dimension $[6.4, 6.9, 4.7]\,\si{m}$, with the exception of a staircase leading to the upper floor of the building, and has an approximate volume of $\SI{208}{\cubic\meter}$. The floor and ceiling are made of wood, and thin line plastered brick walls surround the room. The two shortest walls are each interrupted by two windows of around $\SI{3.3}{\square\meter}$.
On the longest sides there are two wide passages to adjacent rooms, that were closed off by curtains during all recordings. The staircase has a plastered housing with railing made of glass and wooden stairs. \revOne{The room has a reverberation time $\mathrm{T}_{20} = \SI{0.5}{\second}$, estimated according to the ISO 3382-1 standard as detailed in~\cite{dietzenMYRiAD2023}}. The room also has permanent audio equipment installed, but this was not used for this database. Details on the hardware used for the recordings are given in Sec.~\ref{sec:spatial_setup} and Sec.~\ref{sec:mic_configuration}.

\section{Recording equipment}\label{sec:recording_equipment}
A detailed list of the equipment employed for recording the database is provided in Tab.~\ref{tab:list_recording_equipment}. The recording chains used for the moving and RIR recordings are similar. The input signal was sent to the loudspeakers using Adobe Audition for the RIR recordings, and via a Python script for the recordings during motion. The Python script jointly controlled the audio playback and the motion of the robotic cart, as detailed in~\cite{alma9993576358501488}.
In both cases, the loudspeaker signal was sent via USB to the RME Digiface, routed to the RME M-32 DA using the ADAT protocol, and finally sent to the Genelec 8030 CP loudspeakers. Microphone signals were sent to an RME Micstasy, converted to ADAT, and routed to the RME Digiface, where they were acquired using Adobe Audition for the RIR recordings and the Python script for recordings during motion.
In both cases, the acquired signals were stored as audio files in the wav format. The total measured latency of the system was compensated in the RIRs. Both MATLAB and Python scripts were used for post-processing. Details on the cart and rail system used for the recordings during motion are provided in Sec.~\ref{sec:spatial_setup}.

\begin{table*}[ht]
    \caption{List of recording equipment used for creating the database.}
    \label{tab:list_recording_equipment}
    \resizebox{\linewidth}{!}{
    \centering
    \begin{tabular}{lllll}
    \toprule
    \textbf{Type}     &                                   &                                       & \textbf{Product}                 & \textbf{Mic. Config.} \\
    \midrule
    \textbf{Hardware} & \textbf{Reproduction}             & \textbf{Loudspeakers}                 & Genelec 8030 CP                  &                       \\
                      &                                   & \textbf{DA-converters}                & RME M-32 DA                      &                       \\
                      & \textbf{Acquisition}              & \textbf{Microphones}                  & DPA 4090                         & MC1, MC2, MC3            \\
                      &                                   &                                       & AKG CK32                       & MC1, MC2                 \\
                      &                                   &                                       & Neumann KU-100 DH                & MC1                    \\
                      &                                   &                                       & Rode NT-SF1                      & MC3                    \\
                      &                                   & \textbf{AD-converters/pre-amplifiers} & RME Micstasy                     &                       \\
                      & \textbf{Digital interface}        &                                       & RME Digiface USB audio interface &                       \\
                      &                                   &                                       & Apple iMac                       &                       \\
    \textbf{Software} & \textbf{Reproduction/acquisition} &                                       & Adobe Audition                   &                       \\
                      &                                   &                                       & Python                           &                       \\
                      & \textbf{Post-processing}          &                                       & MATLAB                           &                       \\
                      &                                   &                                       & Python                           &                       \\
    \bottomrule
    \end{tabular}
    }
\end{table*}
\section{Spatial setup}\label{sec:spatial_setup}
In this section, we describe the spatial setup of the trajectory and the loudspeakers used to record the database. Throughout the paper, all the provided geometrical information will be relative to the rail system defining the trajectory. The origin of the coordinate system is defined at the start of the trajectory, with the $y$ axis pointing towards the front (i.e., the direction of motion at the start position, see Fig.~\ref{fig:trajectory_scheme}). \secRound{In Fig.~\ref{fig:room_sketch}, we sketch the positioning of the rail within the room and report the room dimensions, the distances of the origin from the walls of the room, and the orientation of the trajectory. This information has been retrieved using measurements of the distances between two reference points on the rail and four reference points on the back wall of the room (the one opposite to the staircase) with known coordinates, exploiting Euclidean distance matrices~\cite{dokmanicEuclideanDistanceMatrices2015}.} The geometry of the system can be visualized using the accompanying code \revOne{described in Sec.~\ref{sec:using_the_database}}.

\subsection{Rail system configuration}
\label{subsec:rail_system_configuration}
The trajectoRIR database relies on the use of a rail system to define a trajectory in the AIL room, along which all (stationary and moving) recordings are performed. The rail system follows the modular design introduced in~\cite{alma9993576358501488}: all rail components are built using Medium-Density Fiberboard (MDF), cut with a thickness of $\SI{6}{\milli\meter}$ using a Trotec Laser Cutter. The track consists of modular blocks that can be assembled to create a custom trajectory shape, along which a robotic cart moves carrying the microphone arrays used for the recordings. The CAD files used to create all the rail and cart blocks are available at~\cite{vanOeterenThesisStadius2023}.

\begin{figure*}
    \centering
    \includegraphics[width=0.7\linewidth]{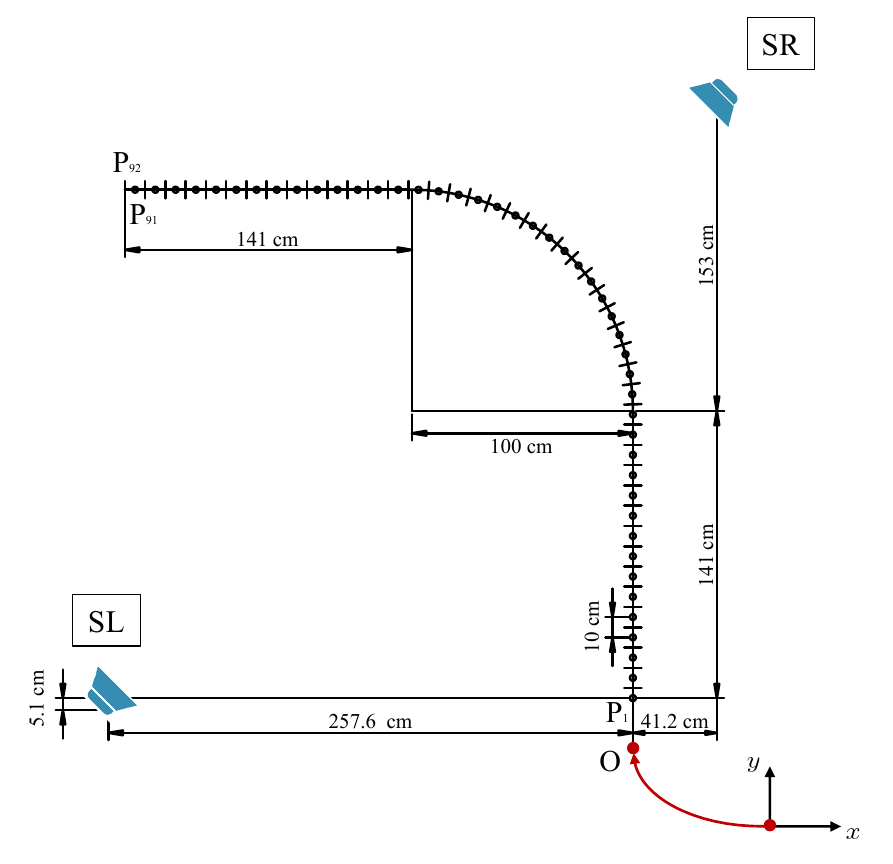}
    \caption{\revOne{Scheme of the trajectory built using the rail system and used to record the database. The direction of the cartesian axes is also reported for reference (the actual coordinate system is centered in P1, as indicated by the red arrow). Loudspeaker positions are labeled SL (loudspeaker left) and SR (loudspeaker right). All indicated dimensions are approximate and only for illustrative purposes: accurate geometrical information is provided in the database.}}
    \label{fig:trajectory_scheme}
\end{figure*}

\begin{figure*}
    \centering
    \includegraphics[width=0.7\linewidth]{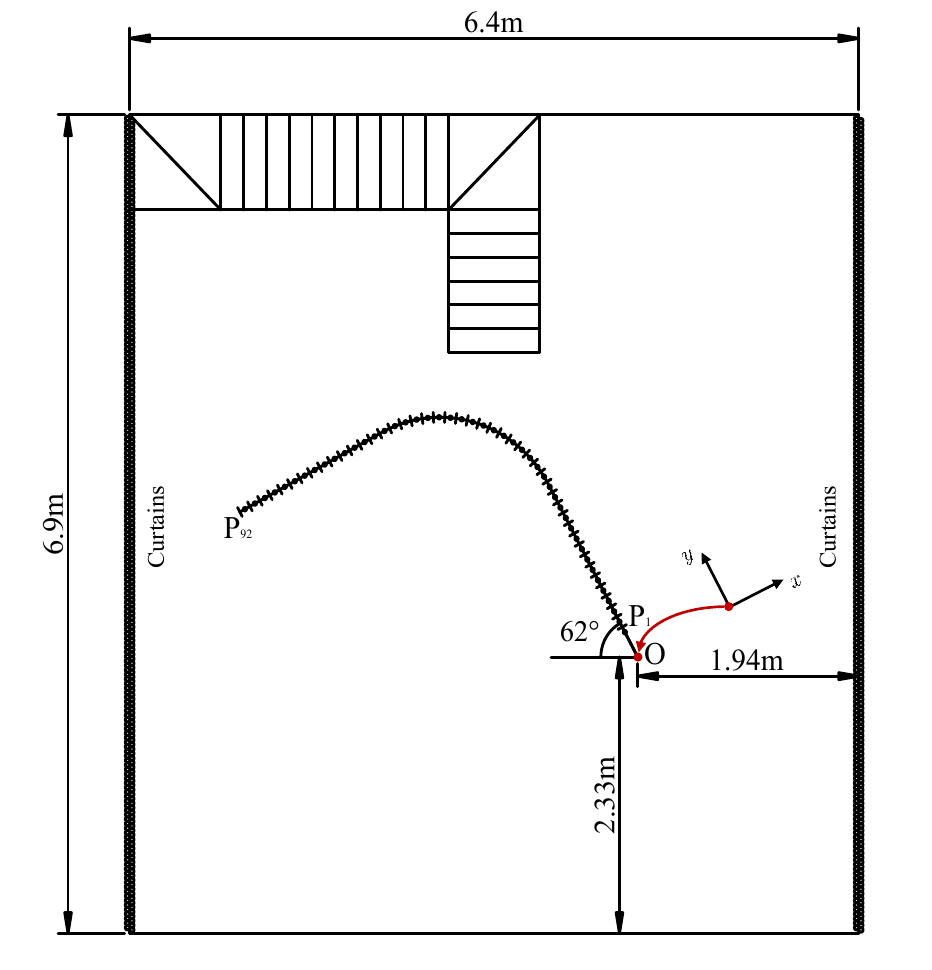}
    \caption{\secRound{Positioning of the rail system within the AIL room. The absolute coordinates of all positions and of the two loudspeakers can be retrieved using the geometrical information provided in the database.}}
    \label{fig:room_sketch}
\end{figure*}

For the trajectoRIR database, we assembled a smooth L-shaped trajectory consisting of two straight segments connected by a curved segment, as depicted in Fig.~\ref{fig:trajectory_scheme}. The trajectory is positioned at the center of the AIL room, with the straight segments oriented such that they are not parallel to any of the walls, \secRound{as depicted in Fig.~\ref{fig:room_sketch}}. The rail system has a width of approximately $\SI{16.95}{\cm}$, and all distances reported in this paper are measured at the outer margin of the outer rail. The trajectory spans from the right end to the left end, covering a total length of approximately $\SI{4.4}{\meter}$ measured at the center of the rail. Note that the total length of the rail is longer, as a margin between the last sampling position and the end of the rail track is required to allow the cart to move.  
The straight segments have a total length of approximately $\SI{1.41}{\meter}$, whereas the curved segment has a radius of approximately $\SI{1}{\meter}$ and covers an angle of $\SI{90}{\degree}$.

The rail is mounted on a series of 10 equally spaced stands, that ensure the stability of the system. The height of the rail and cart system is set to $\SI{1.26}{\m}$. The microphone arrays are designed using MDF supports, mounted on top of the DH in the MC1 configuration, and on a rod in the MC2 and MC3 configurations. To both record RIRs and track the movement of the robotic cart during moving recordings, we manually marked a series of 92 positions along the trajectory, spaced by approximately $\SI{5}{\cm}$. Details of the microphone array setups are provided in Sec.~\ref{sec:mic_configuration}, whereas position and speed labels are provided in Tab.~\ref{tab:pos_speed_labels}.


\begin{figure}
    \centering
    \includegraphics[width=\linewidth]{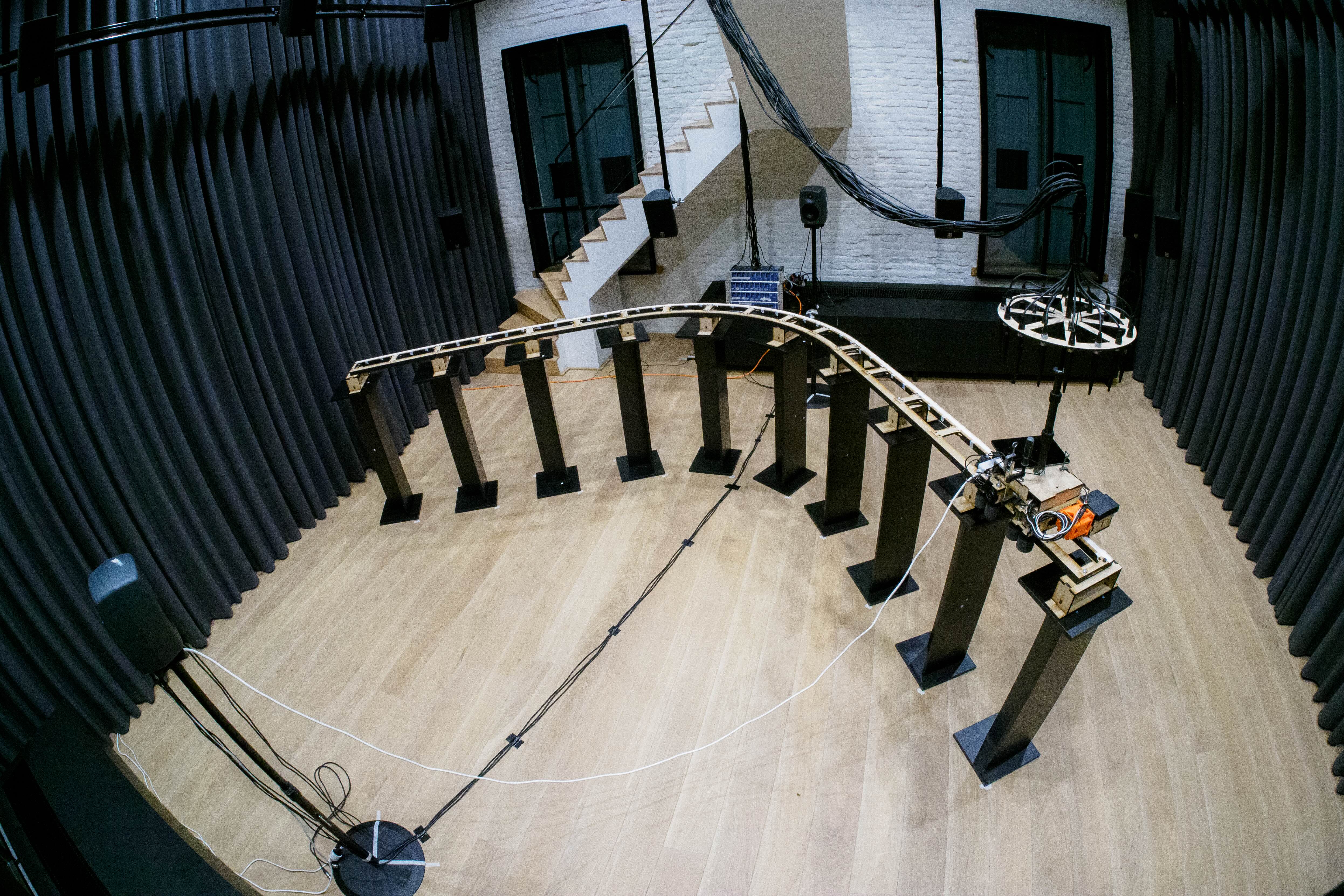}
    \caption{View of the recording setup in the AIL room, with the MC2 array configuration.}
    \label{fig:room_setup_photo}
\end{figure}

\subsection{Loudspeaker configurations}\label{subsec:loudpeaker_configuration}
Two loudspeakers are placed in the AIL room as sound sources for the recordings. They are located at fixed positions on opposite sides of the trajectory and are named according to their respective sides.
The left loudspeaker (SL) and right loudspeaker (SR) are positioned within the inner and outer region of the curve, respectively. \secRound{This positioning ensures a direct path from SL to the left ear of the DH and from SR to the right ear. Moreover, due to the shape of the curve, the microphones move closer to SR while symmetrically moving away from SL along the first rectilinear segment before the curve, and vice versa afterward.}
Both loudspeakers are oriented towards the curved part of the rail trajectory. According to the directivity patterns provided by the manufacturer, a maximum drop of $\SI{5}{\decibel}$ in the direction of the two ends of the trajectory (i.e., P1 and P92) is expected at high frequencies. A scheme of the location of the two loudspeakers with respect to the trajectory is shown in Fig.~\ref{fig:trajectory_scheme}. The position of the two loudspeakers can be extracted using the accompanying code. Both loudspeakers are located at a height of $\SI{1.58}{\meter}$, measured from the floor to the top of the lowest cone.

\section{Microphone configurations}\label{sec:mic_configuration}
In this section, we describe the microphone arrays used for the recordings along the trajectory. All microphone and loudspeaker labels are summarized in Tab.~\ref{tab:labels}, and the coordinates of all microphone positions along the trajectory can be retrieved using the accompanying code, \revOne{as will be detailed in Sec.~\ref{sec:using_the_database}}. The exact positioning of the microphones on the cart has been performed using a crossline laser. All microphone array supports are mounted at the same height of $\SI{1.782}{\meter}$. For MC1, this corresponds to the height from the floor to the top of the DH and resembles the height of an average person. A picture of the mounted system, including the MC2 array configuration, is shown in Fig.~\ref{fig:room_setup_photo}.
\subsection{MC1 and MC2}
The first two microphone configurations have similar characteristics, with the only difference being that the DH is present in MC1 but not in MC2. The configurations contain:
\begin{itemize}
    \item the in-ear microphones of the DH (only for MC1);
    \item two reference (RF) DPA 4090 microphones located in front of the ear canals of the DH and at a horizontal distance of $\SI{1}{\cm}$;
    \item a uniform circular array (UCA) of 16 DPA 4090 microphones, with a radius of $\SI{20}{\cm}$ at the height of the ear canals;
    \item a ``crown array" (CR), consisting of 4 AKG CK32 microphones placed uniformly on a circle with radius $\SI{10.5}{\cm}$, and a height corresponding to the top of the DH.
\end{itemize} 
The two configurations are mounted on an MDF support at the same height, corresponding to the top of the DH in MC1. The DPA 4090 microphones face downwards, with their capsules aligned at the same height as the in-ear microphones of the DH, $\SI{164.8}{\cm}$ above the floor. The AKG CK32 microphones are mounted facing upwards, with the capsules at a height of $\SI{181.6}{\cm}$. 

\begin{figure*}
    \centering
    \begin{subfigure}[tb]{\textwidth}
        \centering
        \raisebox{-0.5\height}{\includegraphics[width=0.37\textwidth]{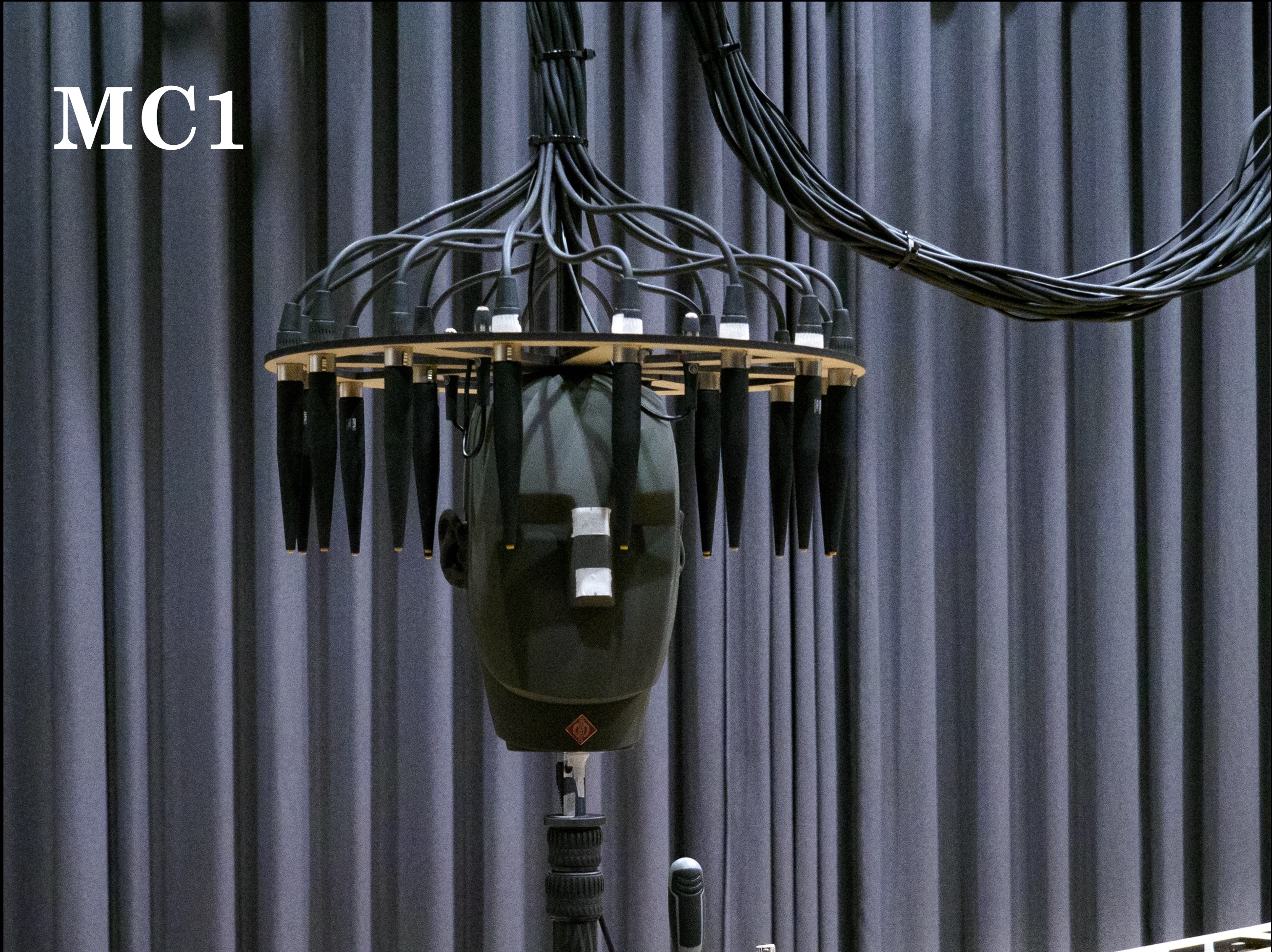}}
        \hspace{2cm}
        \raisebox{-0.5\height}{\includegraphics[width=0.37\textwidth]{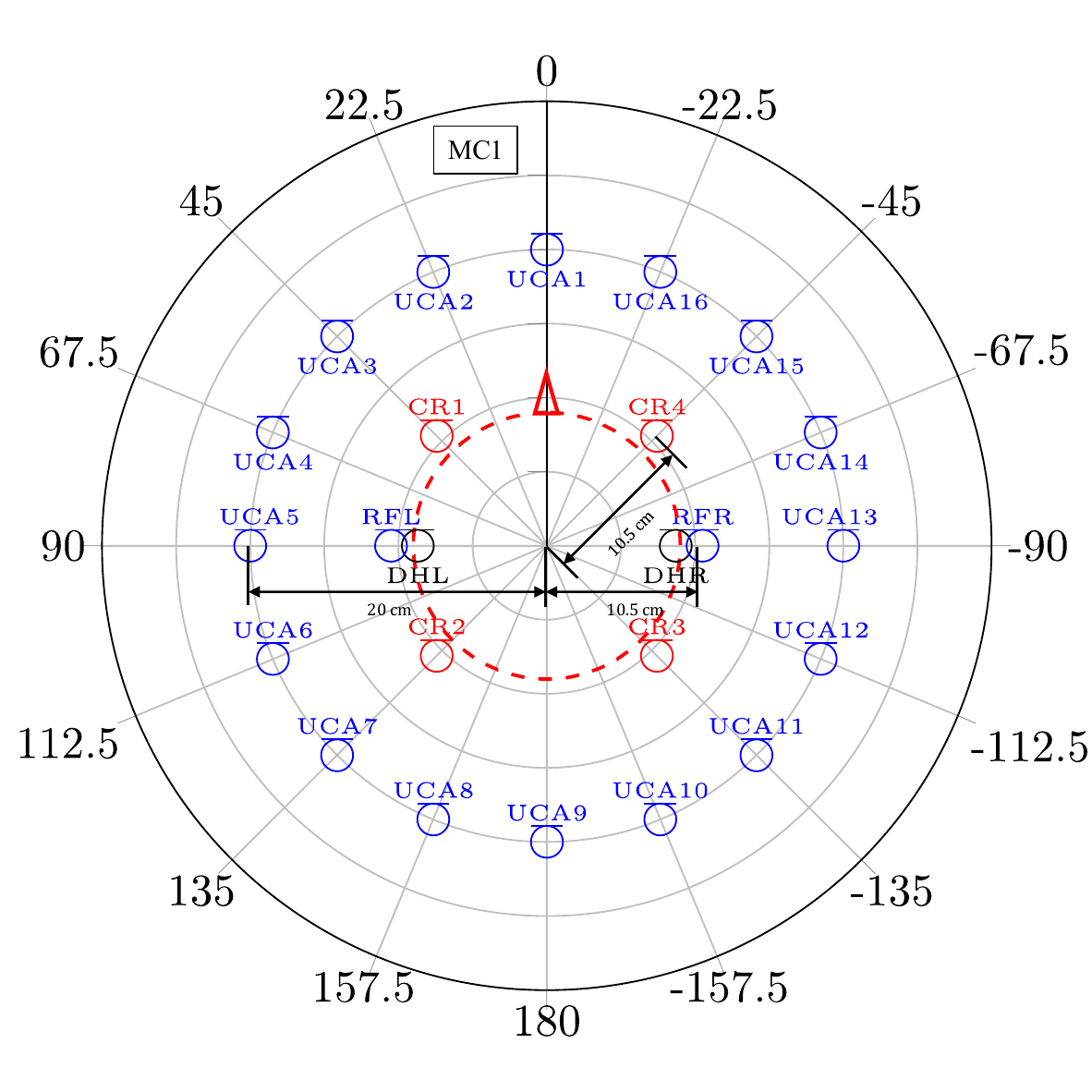}}
        \caption{}
        \label{fig:M1_photo_schematic}
    \end{subfigure}
    
    \begin{subfigure}[tb]{\textwidth}
        \centering
        \raisebox{-0.5\height}{\includegraphics[width=0.37\textwidth]{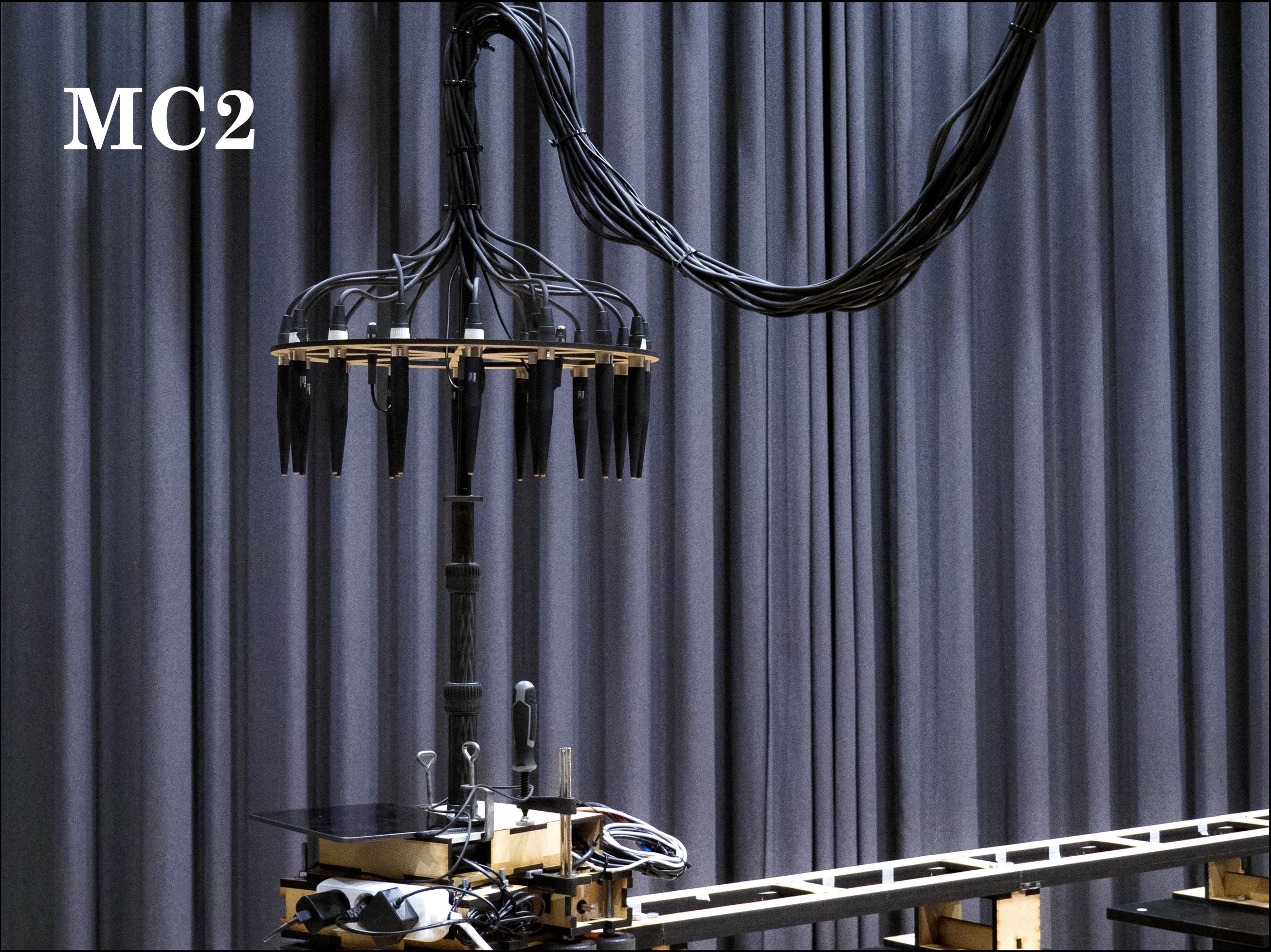}}
        \hspace{2cm}
        \raisebox{-0.5\height}{\includegraphics[width=0.37\textwidth]{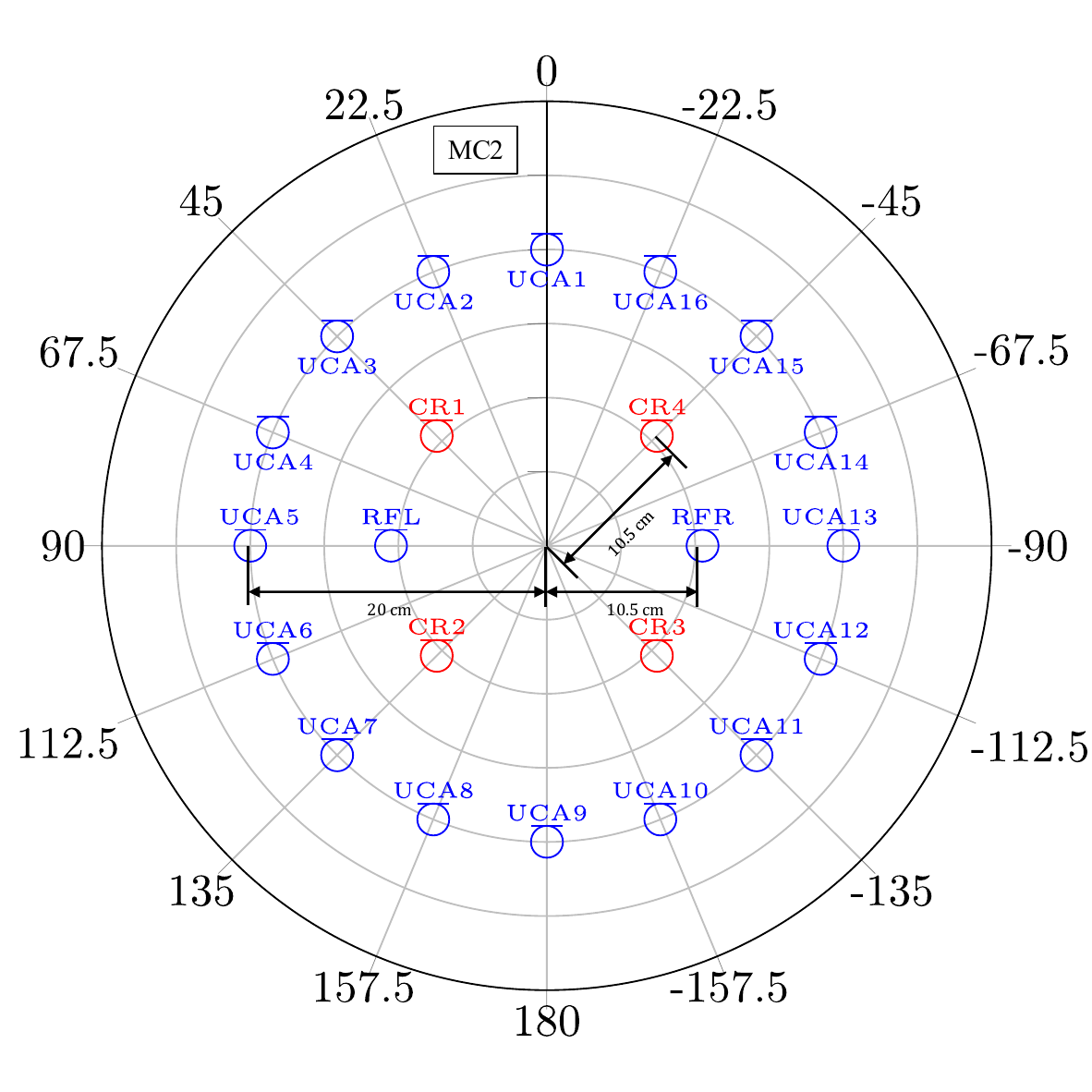}}
        \caption{}
        \label{fig:M2_photo_schematic}
    \end{subfigure}
    
    \begin{subfigure}[tb]{\textwidth}
        \centering
        \raisebox{-0.5\height}{\includegraphics[width=0.37\textwidth]{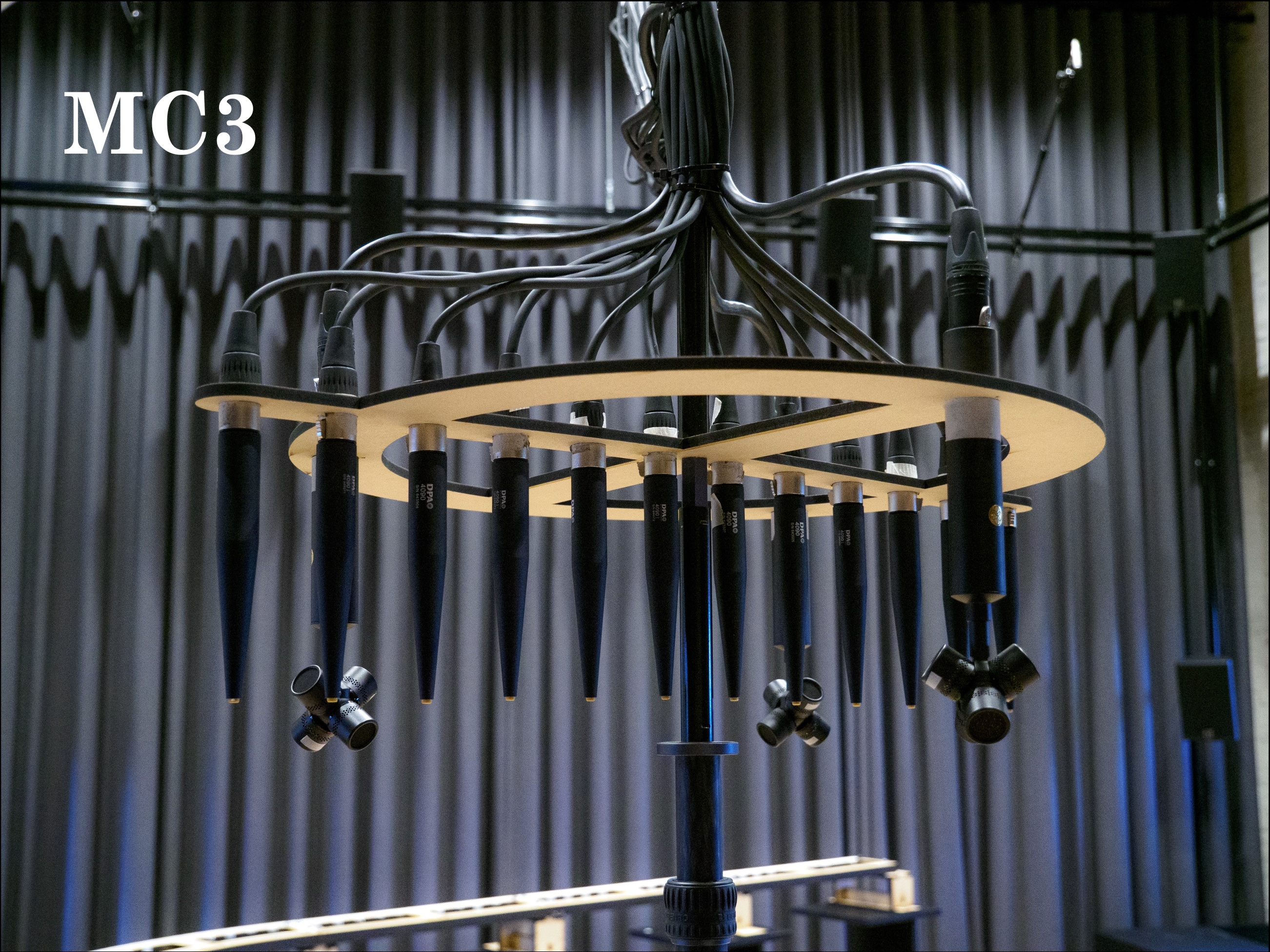}}
        \hspace{2cm}
        \raisebox{-0.5\height}{\includegraphics[width=0.37\textwidth]{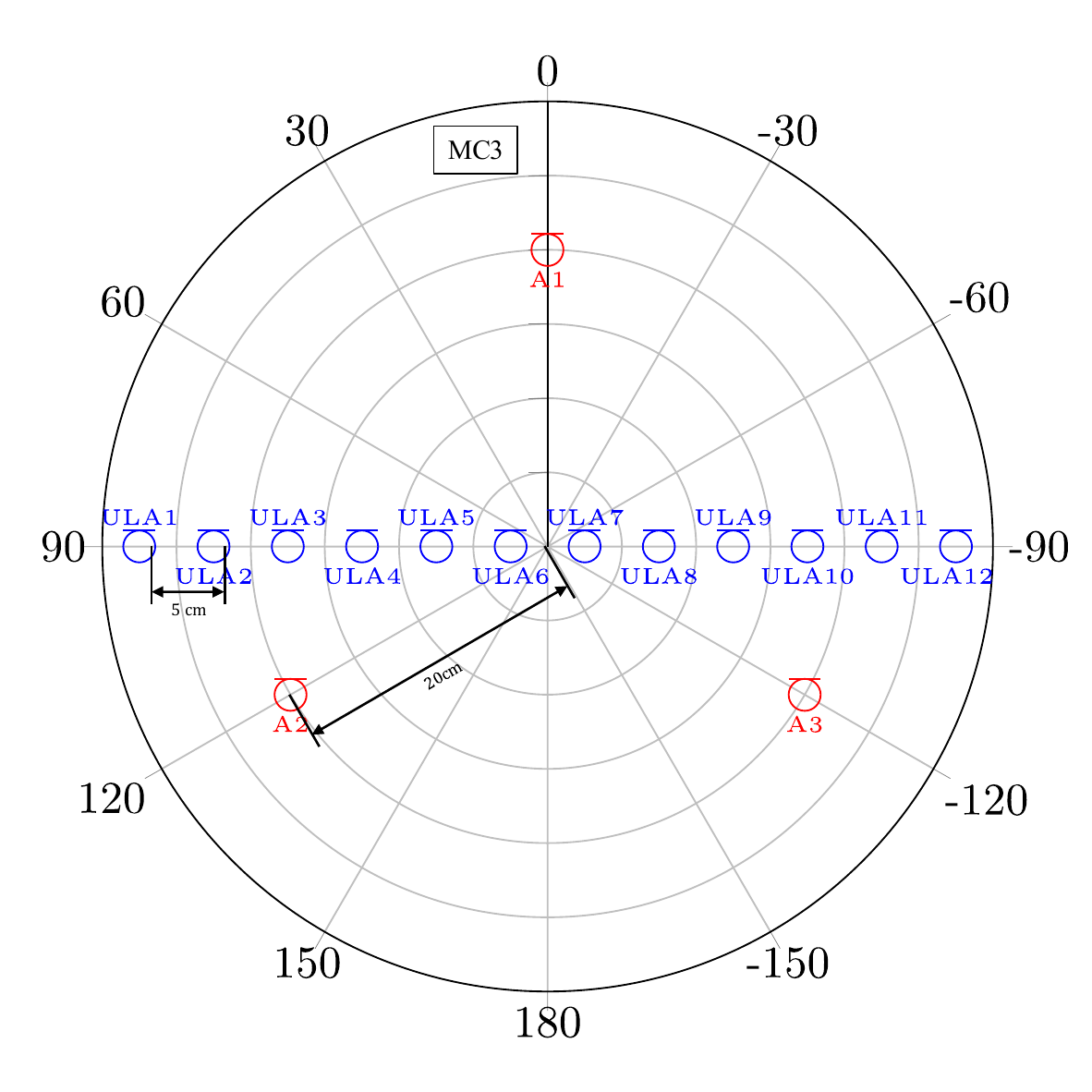}}
        \caption{}
    \label{fig:M3_photo_schematic}
    \end{subfigure}
    
    \caption{Pictures (left column) and top view of the polar plots (right column) of the MC1 (a), MC2 (b) and MC3 (c) microphone array configurations.}
    \label{fig:array_configurations}
\end{figure*}

Fig.~\ref{fig:M1_photo_schematic} and Fig.~\ref{fig:M2_photo_schematic} show the photos and schematics of the two configurations, with the DH facing towards $\SI{0}{\degree}$ (the nose of the DH is marked by a red triangle).

\subsection{MC3}
The third configuration, MC3, consists of a ULA with 12 DPA 4090 microphones, each having an inter-microphone distance of $\SI{5}{\cm}$, and a circular array of three Rode NT-SF1 Ambisonics microphones with a radius of $\SI{20}{\cm}$, as depicted in Fig.~\ref{fig:M3_photo_schematic}.
The configuration is mounted on an MDF support with all microphones facing downwards. The DPA 4090 capsules and the centers of the Rode NT-SF1 microphones are positioned at a height of $\SI{164.8}{\cm}$ above the ground, consistent with MC1 and MC2. The Rode NT-SF1 microphones are oriented radially, with the front microphones facing outwards. Note that the four capsules are identified by their direction (right-front, left-front, right-back, left-back, defined for the microphone being oriented upwards and viewed from the front).

\subsection{Array installation on cart}\label{subsec:array_mounting}
The configurations are mounted on the cart with the direction $\SI{0}{\degree}$ pointing towards the direction of movement. In other words, the schematics plotted in Fig.~\ref{fig:array_configurations} represent relative coordinates, that are translated and rotated with respect to the absolute Cartesian coordinate system during movement. The absolute coordinates of the microphones are available in the metadata, as discussed in Sec.~\ref{subsec:geometry_information}. This results in the DH emulating a person walking along the trajectory, with the left (right) ear on the same side as the left (right) loudspeaker. Due to a horizontal offset between the top and the bottom mounting screws of the dummy head, MC1 is centered $\SI{1.85}{\cm}$ in front of the mounting point on the cart, resulting in a small offset of the RIR measurement positions between the MC1 and MC2 configurations.


\section{Recorded signals}\label{sec:recorded_signals}
The trajectoRIR database contains 3.4 hours of audio recordings, with a total size of 1.25 GB for stationary recordings and 6.22 GB for moving recordings. All recordings were captured at a sampling rate of $\SI{48}{\kilo\hertz}$ with a 24-bit resolution per sample. All recordings were made in the AIL room, where two sides of the room were enclosed with heavy curtains to reduce reflections and external noise. \revTwo{Note that the $T_{20}=\SI{0.5}{\second}$ of the room was estimated from recordings made with the curtains, the effect is therefore already included in the estimate~\cite{dietzenMYRiAD2023}.}

The stationary recordings (STAT) involved capturing RIRs between each loudspeaker (SL and SR) and the various microphone configurations at marked positions along the trajectory. The recordings during motion (MOV) captured six distinct source signals from each loudspeaker while the microphone array configurations moved along the trajectory at three different speeds. To ensure similar signal levels, gains were applied during post-processing to all microphone signals such that the reverberant tails of the measured RIRs were approximately equal across microphones.
Furthermore, system latencies between all recorded and source signals were compensated as detailed in Sec.~\ref{rirs} and Sec.~\ref{playback}.
 
A summary of all recorded and computed signals is provided in Tab.~\ref{tab:signals}. The subsequent sections detail the recording and processing methodologies for both the RIRs and moving microphone signals.
\subsection{Room impulse responses}\label{rirs}
To obtain the RIRs, two exponential sine sweeps were played sequentially by one speaker at a time and recorded using the various microphone configurations at each marked position along the track. Microphone array configurations MC1 and MC2 sampled 46 positions along the trajectory, spaced approximately $\SI{10}{\cm}$ apart, while configuration MC3 sampled 92 positions with a finer spacing of approximately $\SI{5}{\cm}$.
In total, 8648 RIRs were captured across all configurations and loudspeaker-microphone combinations. The playback and recording processes were managed using Adobe Audition, with each microphone in the relevant configuration assigned to a separate channel. 

The recorded sweeps were processed using cross-correlation to compute the RIRs, following the method outlined in \cite{holtersImpulseResponseMeasurement2009}. From the two resulting sets of RIRs computed from the two recorded sweeps, we selected only one for the database according to the following procedure. The sets of RIRs were convolved with the original sweeps, and the set yielding the closest match between the result of this convolution and the recorded sweeps was chosen, with the normalized misalignment serving as the criterion. Additionally, system latency was measured using a loop-back signal and compensated for in the RIRs during post-processing.

\begin{figure}
    \centering
    \includegraphics[width=\linewidth]{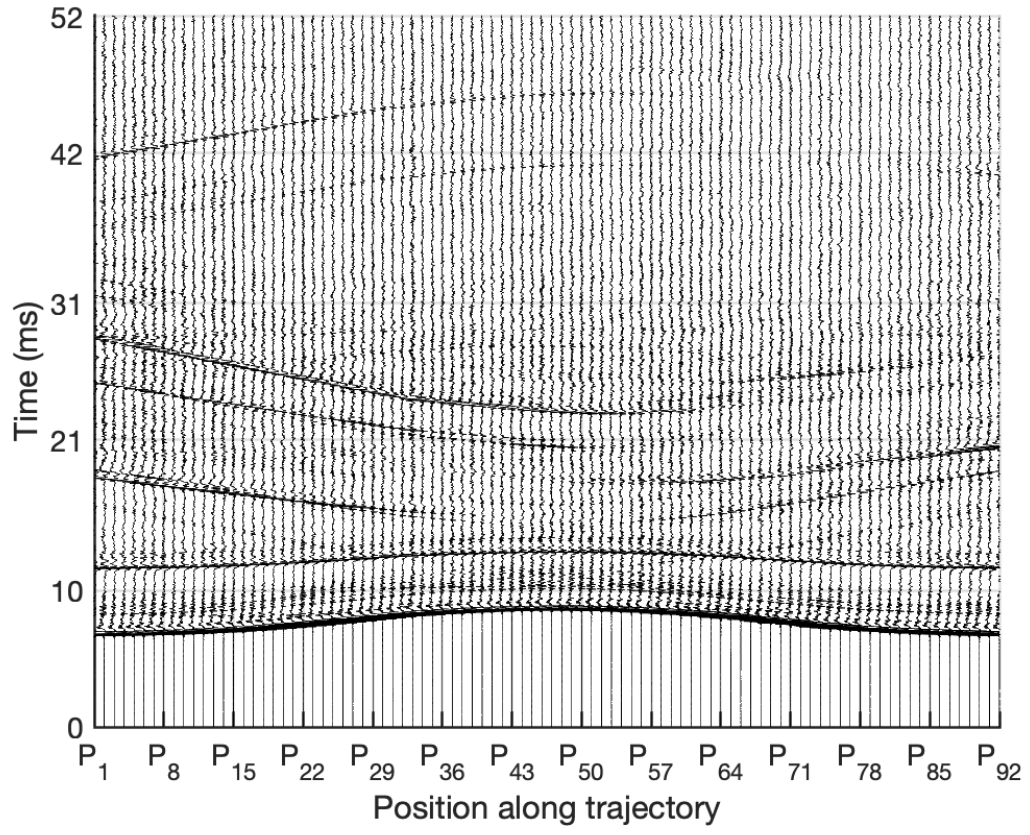}
    \caption{Stacked plot of RIRs along the trajectory, computed using the ULA1 microphone of the MC3 configuration and the SL loudspeaker.}
    \label{fig:RIRstack}
\end{figure}

Fig.~\ref{fig:RIRstack} shows a stacked plot of the RIRs computed at the 92 positions along the trajectory using the ULA1 microphone of the MC3 configuration and SL. The varying distance between the microphone and the loudspeaker is clearly visible from the varying time of arrival of the direct component at the different positions.


\subsection{Recordings during motion}
In the recordings captured during motion, the microphone array configurations traveled along the trajectory illustrated in Fig.~\ref{fig:trajectory_scheme}, moving at three constant speeds: $\SI{0.2}{\meter\per\second}$, $\SI{0.4}{\meter\per\second}$, and $\SI{0.8}{\meter\per\second}$. The movement of the cart was controlled via a Python-based web server as described in \cite{alma9993576358501488}. For each combination of microphone configuration, loudspeaker, and speed, six different source signals were played and recorded. These source signals comprised a piano piece ($\text{PI}$), a drum track ($\text{DR}$), a female speech signal ($\text{SP}$), a white noise signal ($\text{WN}$), and two perfect sweeps \cite{Antweiler2012} ($\text{PS1}$ and $\text{PS8}$) extending up to $\SI{1}{\kilo\Hz}$ and $\SI{8}{\kilo\Hz}$, respectively. 

\begin{figure*}
    \centering  
    \begin{subfigure}[tb]{0.32\textwidth}
        \centering
        \includegraphics[height=0.9\textwidth]{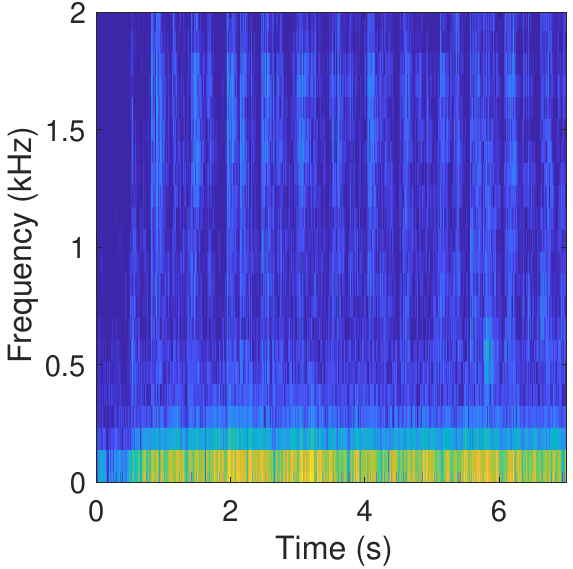}
    \caption{}
    \label{fig:M1_config}
    \end{subfigure}
    \begin{subfigure}[tb]{0.32\textwidth}
        \centering
        \includegraphics[height=0.9\textwidth]{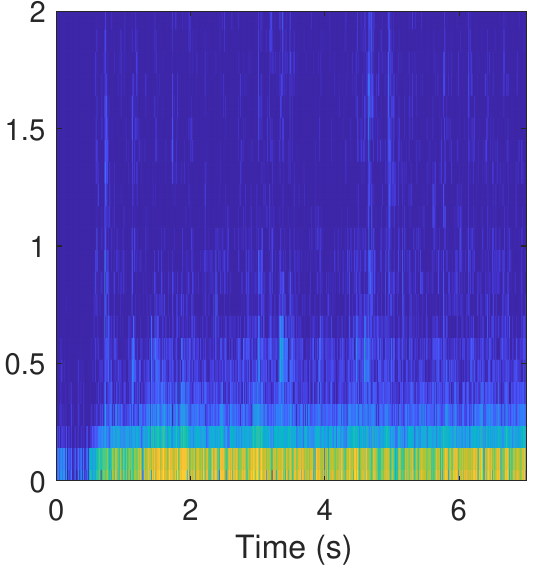}
    \caption{}
    \label{fig:M2_config}
    \end{subfigure}
    \begin{subfigure}[tb]{0.32\textwidth}
        \centering
        \includegraphics[height=0.9\textwidth]{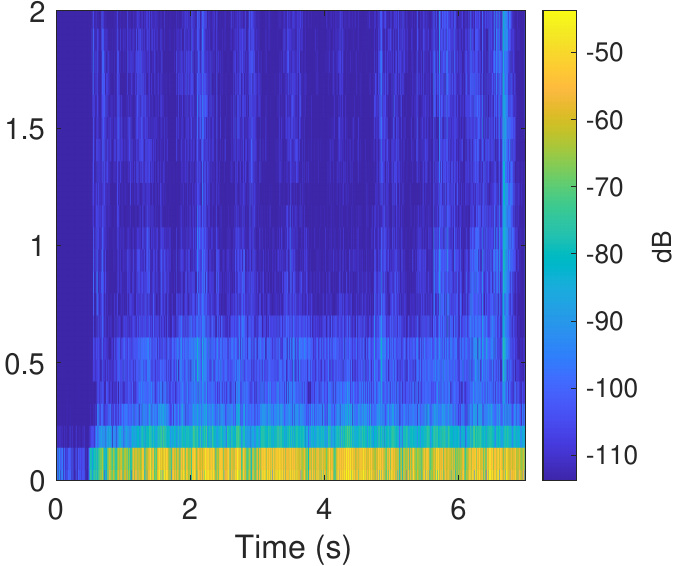}
    \caption{}
    \label{fig:M3_config}
    \end{subfigure}
    \caption{Spectrograms of the `Ego-noise' recording  on microphone ULA1 for MC3 at (a) $\SI{0.2}{\meter\per\second}$, (b) $\SI{0.4}{\meter\per\second}$, and (c) $\SI{0.8}{\meter\per\second}$.}
    \label{fig:silence2}
\end{figure*}

\begin{figure*}[htbp]
    \centering
    \vspace{10pt} 
    \includegraphics[width=\textwidth]{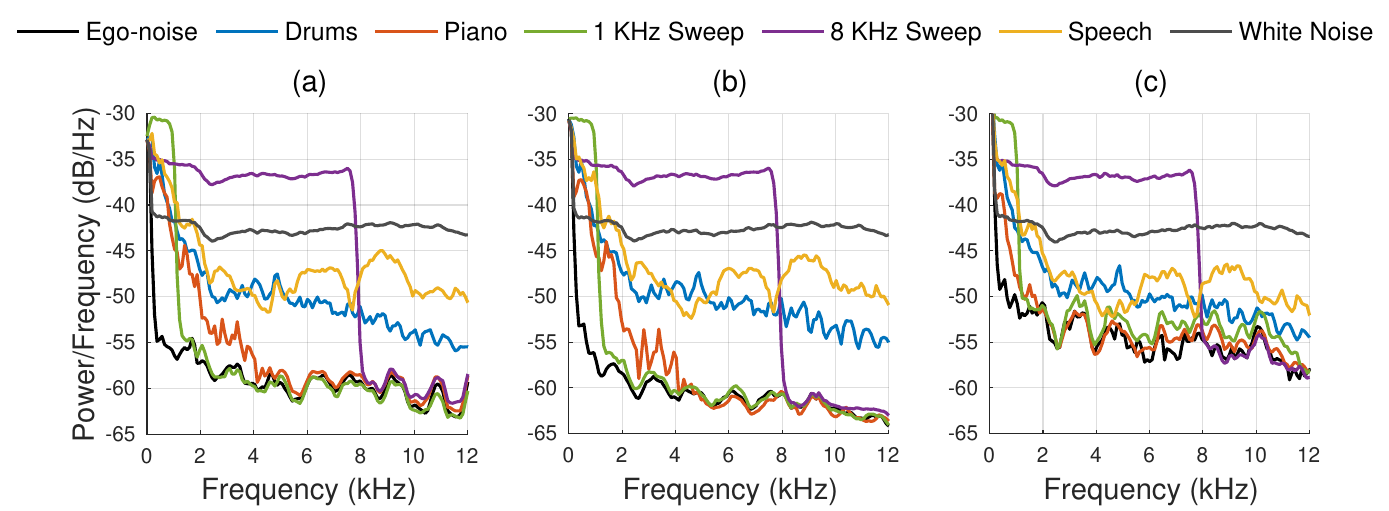}
    \caption{PSD of the various recorded signals on microphone ULA1, using SL and MC3 moving at (a) $\SI{0.2}{\meter\per\second}$, (b) $\SI{0.4}{\meter\per\second}$, and (c) $\SI{0.8}{\meter\per\second}$.}
    \label{fig:PSD}
    \vspace{10pt} 
\end{figure*}

In total, 108 multichannel recordings were made, with 36 recordings for each microphone configuration. To support noise analysis, additional recordings were made capturing only the mechanical noise from the cart’s movement (labeled `Ego-noise’) at the three speeds for the MC3 microphone configuration. Fig.~\ref{fig:silence2} shows spectrograms of the recorded ego-noise on microphone ULA1, highlighting the non-stationarity of the background noise across different speeds.
Fig.~\ref{fig:PSD} displays the Power Spectral Density (PSD) of all recorded signals in MC3, including ego-noise, computed using the Welch method. The results indicate that the majority of the noise energy is concentrated below 250 Hz, and that the recording at $\SI{0.4}{\meter\per\second}$ exhibits the lowest noise floor.

\subsubsection{Measurement position timestamps}
\begin{figure*}[htbp]
    \centering
    \vspace{10pt} 
    \includegraphics[width=\textwidth]{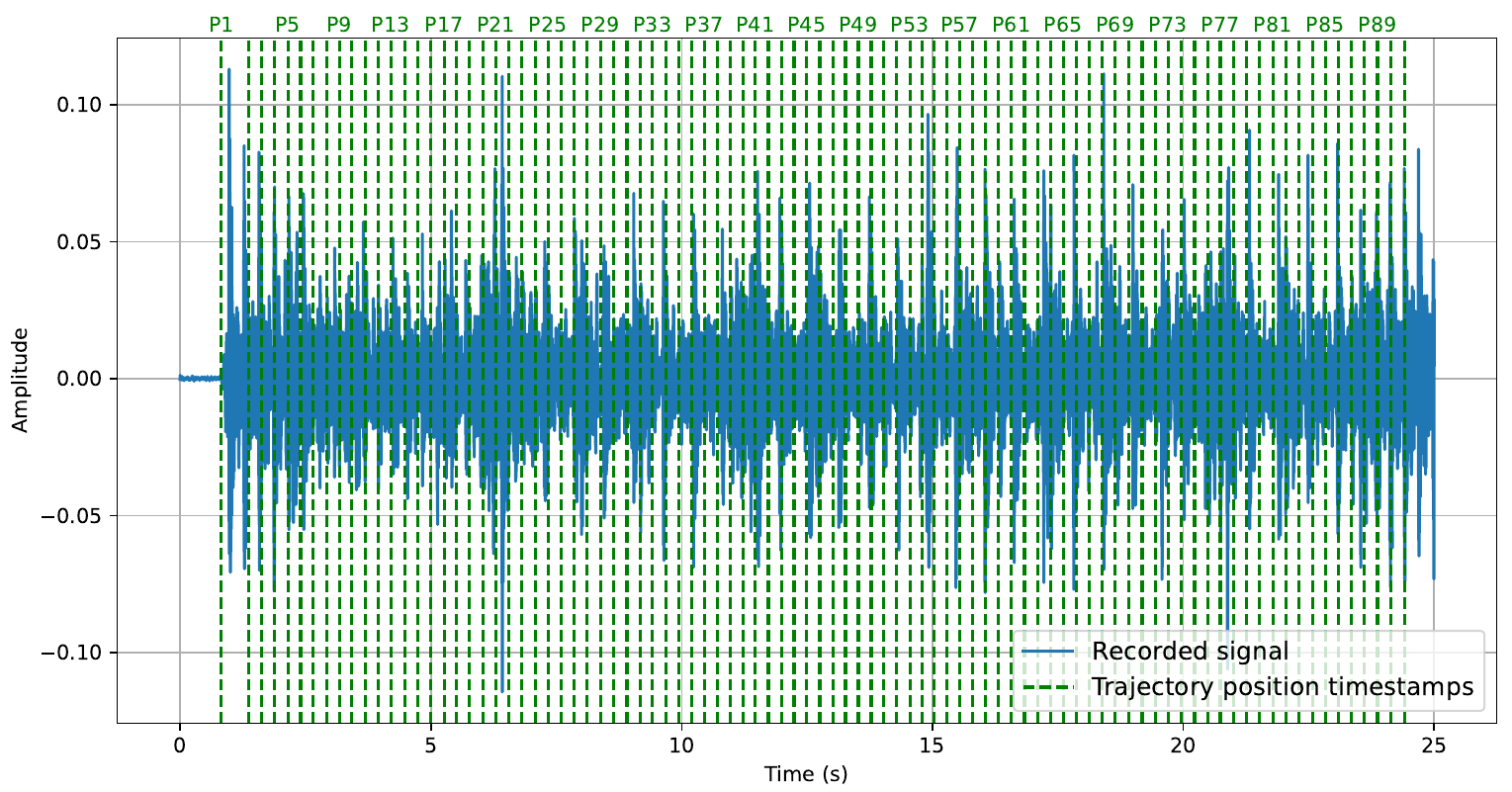}
    \caption{Drum signal from SL recorded using microphone UCA1 in MC1 moving at $0.2$ m/s. The vertical lines indicate the timestamps corresponding to specific positions along the trajectory.}
    \label{fig:timestamps}
    \vspace{10pt} 
\end{figure*}

In order to track the microphone positions along the trajectory in relation to time within the recorded signal, an iPad was mounted on the moving cart to record slow-motion video of the rail below at 240 frames per second (FPS). Using DaVinci Resolve, timestamps were manually added to the footage whenever a fixed reference point on the cart reached an RIR measurement position along the trajectory. The audio track from the video was extracted and a reference microphone was chosen for each microphone configuration (UCA1 for MC1 and MC2, and A1RF for MC3). Cross-correlation was then used to estimate the delay between the audio track from the video and the corresponding reference microphone signal. This delay was used to map the video timestamps and hence the passages of the RIR measurement positions on the timeline of the microphone signals.
The recorded signal was then truncated to end at 25 seconds for $ v = \SI{0.2}{\meter\per\second} $, 15 seconds for $ v = \SI{0.4}{\meter\per\second} $, and 10 seconds for $ v = \SI{0.8}{\meter\per\second} $, well after the cart reached its final position.
Fig.~\ref{fig:timestamps} illustrates a truncated recorded signal overlaid with the trajectory position timestamps. The timestamps are stored in the mov\_timestamps.csv file.



\subsubsection{Latency compensation}\label{playback}
%
The system latency during recordings under motion unexpectedly differed from the measured latency in the RIR recordings. This discrepancy arose from the use of the Python script in \cite{alma9993576358501488} instead of Adobe Audition \revTwo{and resulted in a delay in the start of the microphone recording relative to the intended start time}. As this latency was not measured during the recording session, it had to be estimated from the recorded data to ensure temporal alignment between the recorded signal and the source signal. Because the system latency was found to be consistent across all recordings, we applied a single shift to obtain time-aligned signals, as described in the following paragraphs.

We start by using a variant of the state-space model outlined in \cite{macwilliamStatespaceEstimationSpatially2024} to estimate the so-called time-variant RIR $ h(k,m) $ which relates the source signal, denoted by $x(k) $, to the corresponding recorded signal, denoted by $ y(k) $. Here, $k$ is the discrete-time index and $m $ indexes the time shift of the RIR samples. \secRound{Since RIRs are linear time-invariant (LTI) for any fixed source–receiver location, a ``time-variant'' RIR in this work simply reflects that the microphone’s location changes over time. Thus, a time-variant RIR is effectively a location-variant RIR. Let $l$ denote the microphone’s one-dimensional location index along its trajectory. If the microphone occupies one location per time step $k$, we can set $l = k$, as in \cite{macwilliamStatespaceEstimationSpatially2024}.}

 The model is formulated in state-space form as follows:
\begin{align}
    \mathbf{h}(k) &= \mathbf{h}(k-1) + \mathbf{w}(k), \label{state_equation_prop} \\
    y(k) &= \mathbf{x}^T(k) \mathbf{h}(k) + v(k), \label{observation_equation_prop}
\end{align}
where $ v(k) $ represents a noise term, $ \textbf{w}(k) $ represents process noise modeling errors and
\begin{equation}
\mathbf{x}(k) = \begin{pmatrix} x(k) & x(k-1) & \dots & x(k-M+1) \end{pmatrix}^T,
\end{equation}
\begin{equation}
\mathbf{h}(k) = \begin{pmatrix} h(k,0) & h(k,1) & \dots & h(k,M-1) \end{pmatrix}^T. \label{time-variant_rir}
\end{equation}
An estimate $ \hat{\mathbf{h}}(k) $ of the RIR $ \mathbf{h}(k) $ is then obtained from Eqs.~\eqref{state_equation_prop}-\eqref{observation_equation_prop} using a Kalman filter \cite{simon2006optimal}. \revTwo{Considering that, for alignment purposes, we are primarily interested in obtaining the delay between the direct source components in the estimated and measured RIRs (rather than the detailed structure of later reflections), this method is suitable because the direct path is typically the strongest and most temporally distinct feature in an RIR, and the procedure above reliably captures this peak in the estimated RIR.} This estimation process was carried out only for the white noise signals, as they provided the best and most consistent RIR estimates.

Next, let $ k_{\text{n}} $ denote the time indices corresponding to the timestamps of the RIR measurement positions along the trajectory, where $\text{n}\in \{1,3, 5,\ldots,91\}$ for MC1 and MC2, and $\text{n}\in \{1,2,3,\ldots,92\}$ for MC3. The estimated and measured RIRs at time index $ k_{\text{n}} $ are given by $ \hat{\mathbf{h}}(k_{\text{n}}) $ and $ \mathbf{h}(k_{\text{n}}) $, respectively. The delay $ \tau_{\max} $ corresponding to the maximum normalized cross-correlation (NCC) between the estimated and measured RIRs, $ \hat{\mathbf{h}}(k_{\text{n}}) $ and $ \mathbf{h}(k_{\text{n}}) $, is computed as  
\[
\tau_{\max} = \arg\max_{\tau} \frac{ \sum_{m} h(k_{\text{n}},m)\hat{h}(k_{\text{n}},m-\tau)}{\|\mathbf{h}(k_{\text{n}})\|_2 \|\hat{\mathbf{h}}(k_{\text{n}})\|_2}.
\]

This delay was consistently found to be approximately the same number of samples across all RIR measurement positions, microphone channels, source speeds, and microphone configurations. Consequently, all recorded signals were shifted by this amount to obtain time-aligned signals,
\[
\tilde{y}(k) = y(k - \tau_{\max}).
\]
The aligned signal $\tilde{y}(k)$ is thus synchronized with both the source signal $x(k)$ and the measured RIRs $\mathbf{h}(k_{\mathrm{n}})$.  Note that $ \tilde{y}(k) $ is the version of the recorded signal stored in the database. \revOne{For notational convenience, in the remainder of this paper we still refer to this calibrated version as just $y(k)$}.

\subsection{Temperature recordings}
At the time of each signal recording, the ambient room temperature was documented and is provided in the files temperature\_STAT.csv and temperature\_MOV.csv, corresponding to the RIR and recordings during motion, respectively. \revOne{This information is important because temperature directly affects acoustic properties, and the recordings were captured over several days with no control over room temperature, resulting in observable temperature variations.}
%
Although temperature differences across the entire dataset are expected, it is important that the RIR measurements and recordings during motion associated with the same microphone configuration were captured under similar temperature conditions. In some cases, however, mismatches in temperature were observed within these configuration-specific pairs. %
For example, in the recordings for MC2, the room temperature during the RIR recordings ranged from \(17.2\,^{\circ}\mathrm{C}\) to \(17.9\,^{\circ}\mathrm{C}\), whereas during the moving recordings it ranged from \(19.4\,^{\circ}\mathrm{C}\) to \(20.6\,^{\circ}\mathrm{C}\). Depending on the required level of accuracy, such inconsistencies may need to be corrected and can be done so using the methods described in \cite{postma2016correction, prawda2023time,bhattacharjeeSoundSpeedPerturbation2025}. \revOne{Moreover, temperature fluctuations should be taken into account when trying to reproduce the exact same measurements.}


\section{Using the database}\label{sec:using_the_database}
This section is devoted to discussing how to use the database: specifically, the file path structure is presented in Sec.~\ref{subsec:file_path_structure} and the code provided to retrieve geometrical information and load the signals is introduced in Sec.~\ref{subsec:geometry_information}. 

\begin{table*}[tb]
    \caption{Microphone and loudspeaker labels.}
    \label{tab:labels}
    \resizebox{\linewidth}{!}{
    \centering
    \begin{tabular}{lllll}
    \toprule
         &   \textbf{Mic. Type}                                       & \textbf{Label}                 & \textbf{Description} \\
    \midrule
    \textbf{Microphones} & \textbf{Dummy Head}             & DHL                 & Left Ear  \\
                         &                                 & DHR                 & Right Ear \\
                         & \textbf{Reference Microphones}  & RFL                 & Reference microphone left \\
                         &                                 & RFR                 & Reference microphone right \\
                         & \textbf{Circular microphone array} & UCA[n]           & With index [n] as depicted in Fig.~\ref{fig:M1_photo_schematic}\\ 
                         &                                    &                  & [n] $\in \{1,2,\ldots,16\}$ \\
                         & \textbf{Crown array}               & CR[n]            & With index [n] as depicted in Fig.~\ref{fig:M1_photo_schematic}\\
                         &                                    &                  & [n] $\in \{1,2,\ldots,4\}$ \\ 
                         & \textbf{Linear microphone array} & ULA[n]           & With index [n] as depicted in Fig.~\ref{fig:M3_photo_schematic} \\
                         &                                  &                  & [n] $\in \{1,2,\ldots,12\}$ \\ 
                         & \textbf{Ambisonics}              & A[n]\_[a][b]             & With index [n] as depicted in Fig.~\ref{fig:M3_photo_schematic}\\
                         &                                  &                  & [n] $\in \{1,2,3\}$ ; individual capsules indexed \\ 
                         &                                  &                  & by [a][b], [a] $\in \{\text{L}, \text{R}\}$, [b] $\in \{\text{F}, \text{B}\}$ \\
    \textbf{Loudspeakers} &                                       & SL                   & Left Loudspeaker                      \\
                          &                                       & SR                   & Right Loudspeaker                      \\
    \bottomrule
    \end{tabular}
    }
\end{table*}

\begin{table*}[tb]
    \caption{Signals recorded and computed in the database.}
    \label{tab:signals}
    \resizebox{\linewidth}{!}{
        \centering
        \begin{tabular}{llllll}
        \toprule
           \textbf{Recording set} & \textbf{Signal}& \textbf{Type} & \textbf{Source} & \textbf{Acquisition} & \textbf{Label} \\
         \midrule
          \textbf{STAT} & Sine sweep & Meas. & Generated & Playback + record & - \\
           & RIR & RIR. & Sine sweeps & Computed~\cite{holtersImpulseResponseMeasurement2009} & RIR \\
          \textbf{MOV} & Perfect Sweep - $\SI{1}{\kHz}$ & Meas. & Generated & Playback + record & PS1 \\
             & Perfect Sweep - $\SI{8}{\kHz}$ & Meas. & Generated & Playback + record & PS8 \\
             & White noise & Noise & Generated & Playback + record & WN \\
             & Female speaker & Speech & \cite{yamagishi_CSTRCorpus_2019} & Playback + record & SP \\
             & Drums & Music & \cite{anti-everything_federation_2011} & Playback + record & DR \\
             & Piano & Music & Recorded & Playback + record & PI \\
         \bottomrule
        \end{tabular}
    }
\end{table*}


\begin{table*}[tb]
    \caption{File path structure of the database.}
    \label{tab:file_structure}
    \resizebox{\linewidth}{!}{
        \centering
        \begin{tabular}{llllllll}
        \toprule
          & \textbf{Root} &  & \textbf{Signal$^1$} & & & & \\
         \textbf{Source signal} & /audio/ & SRC/ & [s].wav & \\
         \\
         & \textbf{Root} &  \textbf{Rec set} & \textbf{Signal$^1$} & \textbf{Speaker$^2$} & \textbf{Mic. Conf.$^3$} & \textbf{Pos./Speed$^4$} & \textbf{File name$^2$}\\
         \textbf{Mic signal} & /audio/ & STAT/ & RIR/ & [l]/ & [m]/ & P[n]/ & [u].wav \\
         & & MOV/ & [s]/ & [l]/ & [m]/ & V[a]/ & [u].wav \\
         \\
         & \textbf{Root} & \textbf{File name} & & & & & \\
         \textbf{Metadata file} & /meta/ & temperature\_STAT.csv \\
        &  & temperature\_MOV.csv \\
        &  & mov\_timestamps.csv \\
        &  & cart\_pose.csv \\
        &  & mic\_coordinates.csv \\
        \\
         & \textbf{Root} & \textbf{Language} & \textbf{Function$^5$} & & & &\\
         \textbf{Code file} & /tools/ & Python/ & [f].py \\
         \bottomrule
        \end{tabular}
    }
    \begin{itemize}
        \item[] $^1$ The signal label [s] takes the forms as defined in Tab.~\ref{tab:signals}.
        \item[] $^2$ The speaker labels [l] and the microphone label [u] take the forms as defined in Tab.~\ref{tab:labels}.
        \item[] $^3$ The position [n] and speed information [a] take the form as defined in Tab.~\ref{tab:pos_speed_labels}. 
        \item[] $^4$ The microphone configuration label [m] takes the form as defined in Fig.~\ref{fig:array_configurations}.
        \item[] $^5$ The script or function names [f] take the forms as defined in Tab.~\ref{tab:functions}.
    \end{itemize}
\end{table*}


\begin{table*}[tb]
    \caption{Scripts to load and use the database.}
    \label{tab:functions}
    \centering
    \begin{tabular}{@{} p{0.25\textwidth} p{0.7\textwidth} @{}} 
        \toprule
        \textbf{Function name} & \textbf{Description (more details can be found in the documentation)}\\
        \midrule
        load\_RIR\_data & Load RIRs recorded using a chosen loudspeaker, an arbitrary subset of microphones within a microphone array configuration, and an arbitrary set of positions along the trajectory. Geometry and temperature information are also loaded.\\
        load\_mov\_data &  Load recordings during motion using a chosen loudspeaker, speed, audio signal, and an arbitrary subset of microphones within a microphone array configuration. Timestamps and temperature information are also loaded.\\
        load\_coordinates & Load and optionally plot microphone and loudspeaker coordinates.\\
        \bottomrule
    \end{tabular}
\end{table*}

\begin{table*}[tb]
    \caption{Position and speed labels.}
    \label{tab:pos_speed_labels}
    \centering
    \begin{tabular}{llll} 
        \toprule
        & \textbf{Microphone Config.} & \textbf{Position label} & \textbf{Description}\\
        \textbf{STAT} & MC1, MC2 & P[n] & n $\in \{1,3, 5,\ldots,91\}$\\
        & MC3 & P[n] & n $\in \{1,2,3,\ldots,92\}$\\
        \\
        & \textbf{Speed label} & \textbf{Description} \\
        \textbf{MOV} & V1 & $\SI{0.2}{\meter\per\second}$\\
        & V2 & $\SI{0.4}{\meter\per\second}$\\
        & V3 & $\SI{0.8}{\meter\per\second}$\\
        \bottomrule
    \end{tabular}
\end{table*}

\subsection{File path structure}\label{subsec:file_path_structure}
As illustrated in Tab.~\ref{tab:file_structure}, the files in the database are organized in three root folders: /audio/, containing all the audio files in wav format, /meta/, containing all the metadata in csv format, and /tools/, containing the source code. The audio files are arranged in three subfolders: the loudspeaker signals are stored in the SRC/ directory, the RIR recordings are located in the STAT/ directory and the recordings during motion are contained in the MOV/ directory. The RIR recordings are further divided by loudspeaker (SL or SR), recording configuration (MC1, MC2 and MC3), and position of the array on the trajectory. The recordings during motion, instead, are divided by signal type (according to the labels provided in Tab.~\ref{tab:signals}), loudspeaker, microphone array configuration and speed (V1, V2 and V3). In both cases, the file name encodes the specific microphone within the configuration.

The provided metadata is organized in five tables, stored in the /meta/ directory. The temperature\_STAT.csv and temperature\_MOV.csv tables contain the temperature in Celsius degrees registered at the measurement time for the RIR and recordings during motion, respectively. The mov\_timestamps.csv table contains, for each recording during motion, the timestamps corresponding to the time instants when the moving cart passed on each of the 46 (or 92, for MC3) positions marked on the rail. The cart\_pose.csv table contains the position of the two loudspeakers, and the position of the cart (i.e., the mounting point of the microphone array) as well as its horizontal rotation angle for each of the 92 marked positions. Finally, the mic\_coordinates.csv table contains the Cartesian coordinates of each microphone at each of the marked positions on the rail.

The folder /tools/ contains Python scripts for accessing audio data and retrieving geometrical information from the tables described above. Further details are provided in Sec.~\ref{subsec:geometry_information}, and accompanying examples of code usage are provided as well in the /tools/ directory. 

\subsection{Retrieving geometry information}\label{subsec:geometry_information}

The overall positions of the central axis of the cart, which served as a mounting point for the different microphone configurations, were first measured relative to the Cartesian coordinate system originating at the start of the trajectory (see Fig ~\ref{fig:trajectory_scheme}). In particular, for each position the \textit{pose} of the cart was measured, indicating both its position and a rotation angle with respect to the orientation at the first position of the trajectory.

Relative microphone positions were measured with respect to the center of the array for each configuration. By combining the cart pose with the relative microphone positions through shifting and rotation matrix multiplications, all microphone geometry could be retrieved and is provided in the mic\_coordinates.csv file.
The absolute positions of all microphones for the MC1, MC2, and MC3 configurations can be retrieved through the script load\_coordinates provided in /tools/. The script also provides plotting functionality to visualize the (selected) positions on the trajectory. 
\revOne{\section{Use cases}\label{sec:use_cases}}
As previously mentioned, accurate modeling of acoustic propagation is important for a range of tasks, including auralization, speech processing, and soundfield analysis. For a stationary source and microphone, this requires knowledge of the RIR at the microphone location. When the microphone moves, the relevant model becomes a time-variant RIR $\bh(k)$, \secRound{as defined in Sec. \ref{playback}}, where each time index $k$ corresponds to the RIR at the microphone’s instantaneous location along its trajectory.

\revOne{In practice, measuring an RIR at every time step is infeasible. As a result, existing approaches generally estimate $\bh(k)$ using one of the following three types of information: 
(i) a sparse set of stationary RIR measurements,  
(ii) a moving-microphone recording, or  
(iii) a hybrid combination of both.  
This section presents representative use cases for each scenario and illustrates how the trajectoRIR dataset supports their evaluation. Such approaches require calibrated timestamps of the RIR measurement positions to associate the microphone location with the correct sample index in the recorded signal.}

\revOne{All examples in this section use downsampled $16$\,kHz audio from the MC3 microphone configuration, with the cart moving at a constant velocity of $0.8$\,m/s. Under this constant-speed condition, each discrete time index corresponds to an equally spaced spatial sample. The time-variant RIR vector $\bh(k)$ is defined as in Eq.~\eqref{time-variant_rir} and measured RIRs are available at indices $k_n$, for $n=1,\dots,92$, and denoted by $\bh(k_n)$.}

\revOne{For the purpose of this section, we focus exclusively on the early part of each RIR, using truncated RIRs that contain only the early reflections. This is the part most relevant for applications such as speech processing and motion-aware acoustic modeling.}

\revOne{The remainder of this section is organized as follows. The individual estimation approaches are first briefly reviewed in Secs.~\ref{sec:usecase_interpolation} and~\ref{sec:usecase_early_rirs}, after which they are evaluated using the trajectoRIR database in Sec.~\ref{evaluation}.}

\revOne{\subsection{Estimation of time-variant RIRs from sparse RIR measurements}
\label{sec:usecase_interpolation}}
\revOne{The first scenario addresses estimation of \(\bh(k)\) when only sparsely sampled RIRs along the trajectory are known. A standard approach is linear interpolation between consecutive measured RIRs \(\bh(k_n)\) and \(\bh(k_{n+1})\). To ensure consistent alignment, we use dynamic-time-warping (DTW) based matching~\cite{kearney2009dynamic} to identify the time indices of the direct component and prominent reflections in each pair of consecutive RIRs. Denote these as \(\bm_r(k_n)\) and \(\bm_r(k_{n+1})\). The interpolated time indices at any \(k\in[k_n,k_{n+1}]\) are then computed as
\begin{equation}
\bm_r(k)
=
\bm_r(k_n)
+
\Bigl\lfloor 
\beta(k)\,(\bm_r(k_{n+1}) - \bm_r(k_{n})) 
\Bigr\rceil.
\label{sample_based_interpolation}
\end{equation}
where \(\beta(k)\in[0,1]\) satisfies \(\beta(k_n)=0\) and \(\beta(k_{n+1})=1\), and is computed from the relative direct source–microphone distances. Reflection amplitudes are interpolated analogously, yielding the estimate \(\hat{\bh}(k)\) for all intermediate time indices between measured RIR positions.}

\revOne{\subsection{Estimation of time-variant RIRs using moving-microphone signals}
\label{sec:usecase_early_rirs}}
\revOne{We next consider approaches formulated in a state-space representation, where the observation equation relates the unknown time-variant RIR to the recorded microphone signal and the state equation models how $\bh(k)$ evolves over time. Two cases are examined:  
(i) a purely data-driven method that relies solely on the moving-microphone recording, and  
(ii) a hybrid method that uses the microphone recording together with a physical model of the evolution of the direct path and early reflections. The parameters of this physical model are extracted from sparsely available RIR measurements.}

\revOne{\subsubsection{Purely data-driven Kalman-filter estimation}\label{datadriven}}
\revOne{When no RIR measurements are available, the time-variant RIR can be estimated from the known source and moving-microphone signals \((x(k),y(k))\). We adopt the state-space model in Eqs.~\eqref{state_equation_prop}–\eqref{observation_equation_prop},
where the state equation \eqref{state_equation_prop} is a first order Markov process of the form \(\bh(k)=\alpha\bh(k-1)+\bw(k)\) with \(\alpha=1\).
All temporal variability is absorbed into the process noise \(\bw(k)\), yielding a fully data-driven formulation without structural priors. A Kalman filter is applied to obtain the adaptive estimate \(\hat{\bh}(k)\). This approach is closely related to adaptive echo-path tracking \cite{enzner2010bayesian} and can reliably estimate early RIR components even without physical modeling included.}

\revOne{\subsubsection{Hybrid Kalman-filter estimation with sparse RIR measurements}\label{hybrid}}
\revOne{We now incorporate a physical model into the state equation by defining a time-variant transition matrix \(\bA(k) \). This would lead to replacing the state equation in \eqref{state_equation_prop} with 
\begin{align}
    \bh(k) &= \bA(k) \bh(k-1) + \bw(k),
\end{align}  
To this end, we follow the method proposed in \cite{macwilliam2025tracking} and let the trajectory be partitioned into linear segments \(s=1,\dots,S\), each spanning
\begin{equation}
\mathcal{K}_s=\{k_{s|\mathrm{st}},\dots,k_{s|\mathrm{en}}\}.
\end{equation}\
where \( k_{s|\text{st}} \) and \( k_{s|\text{en}} \) are, respectively, the start and end time indices of the segment and correspond to measured RIR positions. Within each segment, it is assumed that individual reflection TOA intervals do not overlap. For \(k\in\mathcal{K}_s\setminus\{k_{s|\mathrm{st}}\}\), we aim to approximate a fixed transition matrix, $\bA_s$, for that segment such that
\begin{equation}
\bh(k)\approx\bA_s\bh(k-1),
\end{equation}
leading to the piecewise-constant transition model
\begin{equation}
\bA(k)=\bA_s,\qquad k\in\mathcal{K}_s\setminus\{k_{s|\mathrm{st}}\}.
\end{equation}
The transition matrices \(\bA_s\), serve as a room acoustic prior assuming sound propagation according to the image source model (ISM) and are estimated by interpolating TOAs of the direct and reflected components between the measured RIRs available at the segment boundaries as detailed in ~\cite{macwilliamStatespaceEstimationSpatially2024,macwilliam2025tracking}. This method therefore requires both a measured moving-microphone signal and calibrated sparse RIR measurements. As in \ref{datadriven}, a Kalman filter is used to obtain an estimate $\hat{\bh}(k)$.}

\revOne{\subsection{Evaluation of estimated time-variant RIRs}\label{evaluation}
We now evaluate the three methods introduced above:  
(1) Linear interpolation (algorithm \textbf{LI}) as in Sec.~\ref{sec:usecase_interpolation},  
(2) the purely data-driven Kalman filter (algorithm  $\textbf{KF-}\alpha$) as in Sec.~\ref{datadriven}, and  
(3) the hybrid Kalman filter (algorithm  $\textbf{KF-}\bA(l)$) as in \ref{hybrid}.  
Two criteria are used:  
(i) accuracy of synthesized moving-microphone signals, and  
(ii) agreement between estimated and measured RIRs at known measurement position indices \(k_n\) that were not used during estimation.}

The synthesized moving-microphone signal $\hat{y}(k)$ is obtained using the respective estimated time-variant RIR $\hat{\mathbf{h}}(k)$ as
\begin{equation}
\hat{y}(k) = \mathbf{x}^T(k) \hat{\mathbf{h}}(k).
\end{equation}
and accuracy is quantified via maximum normalized cross-correlation between \(\hat{y}(k)\) and \(y(k)\), restricted to a lag of \(\pm1\) samples to compensate for minor timing shifts. \secRound{The resulting correlation coefficient is defined as
\begin{equation}
\rho = \max_{\tau \in \{-1,0,1\}}
\frac{\sum_k (\hat{y}(k+\tau)-\overline{\hat{y}}_\tau)(y(k)-\bar{y})}
{\sqrt{\sum_k (\hat{y}(k+\tau)-\overline{\hat{y}}_\tau)^2 \sum_k (y(k)-\bar{y})^2}},
\label{eq:maxcorr_single}
\end{equation}
where $\bar{y}$ is the mean of the measured signal $y(k)$ and $\overline{\hat{y}}_\tau$ is the mean of the shifted estimate $\hat{y}(k+\tau)$.}

To assess individual RIR accuracy, local time alignment is first obtained by selecting the lag that maximizes the normalized cross-correlation:
\begin{equation}
\lambda_{\max}(k_n)
= \arg\max_{\lambda\in\{-1,0,1\}}
\frac{
\bh(k_n)\cdot\hat{\bh}(k_n-\lambda)
}{
\|\bh(k_n)\|_2\|\hat{\bh}(k_n-\lambda)\|_2
}.
\end{equation}

The normalized misalignment \secRound{(NM)} is then computed as
\begin{equation}
\secRound{\mathrm{NM}_{\mathrm{dB}}(k_n)} = 20 \log_{10} \Biggl(
\frac{\|\hat{\boldsymbol{h}}(k_n-\lambda_{\max}(k_n)) - \boldsymbol{h}(k_n)\|_2}
{\|\boldsymbol{h}(k_n)\|_2}
\Biggr),
\label{NM}
\end{equation}
\revOne{\subsubsection{Experiment 1}
The first experiment examines how the spacing of RIR measurement positions affects the accuracy of synthesized moving-microphone signals in the context of the interpolation technique described in \ref{sec:usecase_interpolation}. Using the linear segment \(n=1\)–\(31\), we apply algorithm \textbf{LI} for six different excitation signals (WN, SP, PI, DR, PS8, PS1) and eight RIR measurement spacings $\{5,10,15,25,30,50,75,150\}\text{ cm}.$}

\begin{figure}
    \centering
    \includegraphics[width=\linewidth]{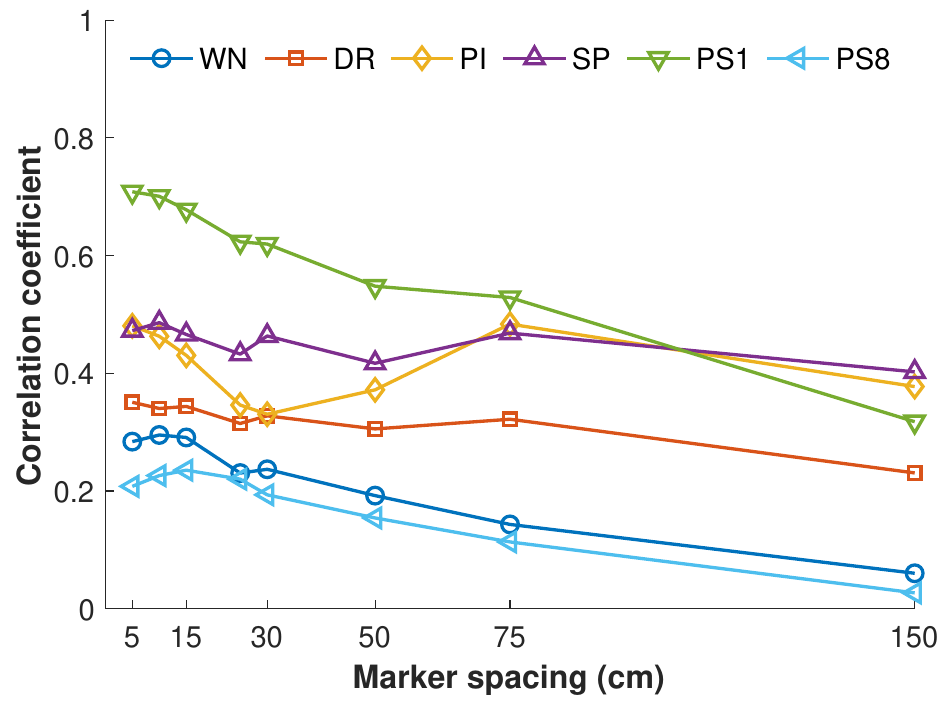}
    \caption{Correlation coefficient \secRound{computed using Eq.~\eqref{eq:maxcorr_single}} between measured and synthesized moving-microphone signals using algorithm $\textbf{LI}$.}
    \label{fig:interp_corr_placeholder}
\end{figure}

Fig.~\ref{fig:interp_corr_placeholder} shows the resulting correlation coefficients for each different spacing used, \secRound{obtained according to Eq.~\eqref{eq:maxcorr_single}.}
Overall, it can be observed that for all excitation signals, the correlation coefficient tends to decrease as the spacing between known RIRs increases, and generally remains relatively low. This indicates that interpolation from sparse RIR measurements alone does not adequately reproduce moving-microphone signals and that interpolation accuracy reduces if spacing increases. Physical effects that are not captured by stationary RIR measurements, including mechanical noise, also contribute to the recorded signal.
\revOne{\subsubsection{Experiment 2}
Since interpolation yields limited accuracy in moving-microphone signal synthesis, as seen in \textbf{Experiment 1}, we now additionally consider the two methods that make use of a moving-microphone recording. Specifically, here the recorded WN signal is used as the observation signal in Eq.~\eqref{observation_equation_prop} for the algorithms $\textbf{KF-}\alpha$ and $\textbf{KF-}\bA(l)$, while the interpolation method $\textbf{LI}$ is still included for comparison.
For simplicity, we focus on the linear section of the full curved trajectory from $k_1$ to $k_{34}$, although the same methods can be applied to the entire trajectory, as demonstrated in~\cite{macwilliam2025tracking}.
Both $\textbf{LI}$ and $\textbf{KF-}\bA(l)$ require sparsely measured RIRs as an input. In this experiment, we use the RIRs at the subset of measurement position indices $\bar{n}= [1,2,3,5,9,13,17,21,23,25,28,32,34]$. Algorithm $\textbf{KF-}\alpha$ relies on the moving-microphone recording alone. Both Kalman filter–based methods are initialized with $\bh(k_1)$.}

\revOne{For all methods, the normalized misalignment defined in Eq.~\eqref{NM} is computed at measurement indices not used during estimation (i.e., excluding $\bar{n}$), and the results are shown in Fig.~\ref{fig:misalignment_placeholder}.
\begin{figure}[t]
    \centering
\includegraphics[width=\linewidth]{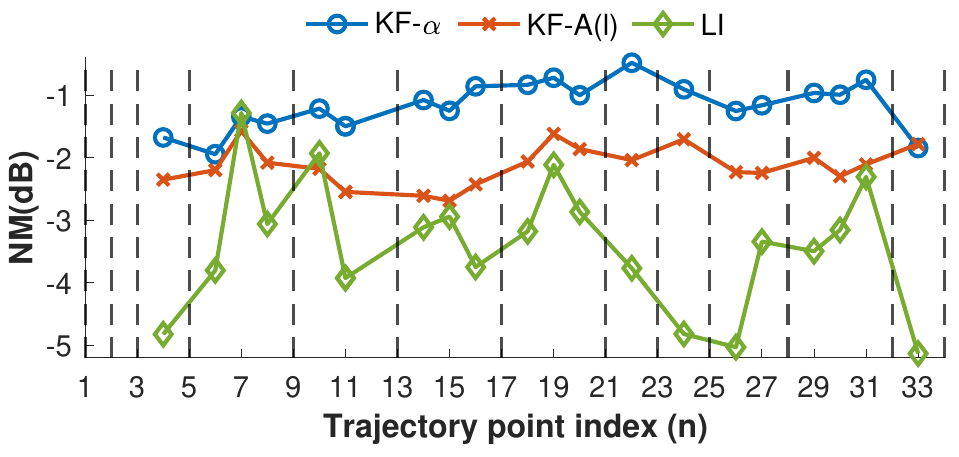}
    \caption{Normalized misalignment between estimated and measured RIRs \secRound{computed according to Eq.~\eqref{NM}. The vertical dashed lines indicate the measurement positions of the RIRs used in $\textbf{LI}$ and $\textbf{KF-}\bA(l)$.}}
    \label{fig:misalignment_placeholder}
\end{figure}
Tab.~\ref{tab:mm_correlation_placeholder} presents the correlation coefficient between synthesized and measured moving-microphone signals for each method and excitation signal, \secRound{computed using Eq.~\eqref{eq:maxcorr_single}}.
\begin{table}[t]
    \centering
    \caption{Correlation coefficient \secRound{computed using Eq.~\eqref{eq:maxcorr_single}} between synthesized and measured moving-microphone signals}
    \label{tab:mm_correlation_placeholder}
    \begin{tabular}{lccc}
        \hline
        & $\textbf{KF-}\alpha$ & $\textbf{KF-}\bA(l)$ & \textbf{LI} \\
        \hline
        WN  & 0.9306  & 0.80724 & 0.58673 \\
        DR  & 0.86285 & 0.76470 & 0.38405 \\
        PI  & 0.59173 & 0.59080 & 0.48307 \\
        SP  & 0.73884 & 0.73793 & 0.52498 \\
        PS1 & 0.83686 & 0.78301 & 0.70912 \\
        PS8 & 0.69060 & 0.61251 & 0.51761 \\
        \hline
    \end{tabular}
\end{table}
It can be seen in Fig.~\ref{fig:misalignment_placeholder} that the algorithm \(\textbf{LI}\) generates time-variant RIRs that most closely match the known stationary measurements, yet as shown in Tab.~\ref{tab:mm_correlation_placeholder} it yields the poorest synthesized microphone signals.  
Conversely, algorithm $\textbf{KF-}\alpha$ provides the most accurate synthesized signals while producing RIR estimates that deviate the most from the stationary measurements.  
The hybrid model, algorithm $\textbf{KF-}\bA(l)$, offers a middle ground, achieving synthesized microphone signals that are almost as well correlated as those produced by $\textbf{KF-}\alpha$, while providing more accurate estimates at the stationary RIR positions. This results in the most consistent performance across both evaluation metrics.}

\revOne{\subsection{Use cases summary}
Overall, the results highlight the importance of jointly using moving-microphone recordings and sparse stationary RIRs for reliable time-variant RIR estimation. The dataset thus allows testing of algorithms aimed at dynamic auralization, acoustic modeling, and audio rendering in motion-dependent scenarios.}

\section{Conclusion}\label{sec:conclusion}
In this paper, we introduced trajectoRIR, a database of room acoustics recordings obtained with microphones moving on a controlled trajectory in a room. The database contains both RIRs, obtained at a set of equally spaced positions along the trajectory, and audio recordings during motion, obtained with the microphones moving along a trajectory at a constant speed. In both stationary recordings and recordings during motion, two loudspeakers were used, located on opposite sides of the trajectory and positioned at equal distances from the start and end points. Three microphone array configurations were used, including two circular arrays, a linear array, a dummy head with two additional reference microphones positioned next to the ear canals, and three first-order ambisonics microphones. For recordings during motion, three speeds were used, as well as six different signals, including sweeps, speech, noise, and music. Python scripts for accessing audio data, coordinates, and temperature are provided and described. 

\secRound{The collection of matched stationary RIRs and recordings of audio during motion along the same controlled trajectory promotes the adoption of trajectoRIR for applications including time-variant RIR estimation, spatially dynamic sound field reconstruction, auralization, and evaluation of dynamic audio simulations. Example use-cases related to time-variant RIR estimation have been discussed in Sec.~\ref{sec:use_cases}.} 

\secRound{Currently, only a single room and geometric setup has been considered. However, the proposed measurement setup is reproducible and, thanks to the modularity of the rail system, additional data with different geometrical configurations (e.g., room, trajectory, microphone and loudspeaker configurations) could be collected in the future, to extend this database.} The database is available at~\cite{trajectoRIR2024}.

\section*{Declarations}

\subsection*{Funding} This research work was carried out at the ESAT Laboratory of KU Leuven, in the frame of KU Leuven internal funds C3/23/056 and C14/21/075, FWO SBO Project S005525N, and FWO Research Project G0A0424N. The research leading to these results has received funding from the Flemish Government (AI Research Program), from the European Union's Horizon 2020 research and innovation programme under the Marie Sk\text{\l}odowska-Curie grant agreements No. 956369 and No. 956962, and from the European Research Council under the European Union's Horizon 2020 research and innovation program / ERC Consolidator Grant: SONORA (no. 773268). This paper reflects only the authors' views and the Union is not liable for any use that may be made of the contained information.

\subsection*{Acknowledgments} The authors would like to thank Dante Van Oeteren and Koen Eelen for their Master's thesis work, that lead to the design of the robotic cart and rail system that enabled the collection of the database presented in this paper.

\subsection*{Authors' contributions} SD, KM, VL, TD and TVW jointly designed the recording setup and methodology. SD, KM, VL and TD acquired and post-processed the audio data, and compiled the database. SD and KM drafted the manuscript. All authors read and reviewed the final manuscript.

\subsection*{Availability of data and materials} The database, including metadata and accompanying code is publicly available at~\cite{trajectoRIR2024}. The CAD files used to create all the rail and cart blocks, as well as the Python code to operate the robotic cart is publicly available at~\cite{vanOeterenThesisStadius2023}.

\subsection*{Conflict of interest} The authors declare that they have no competing interests.

\subsection*{List of Abbreviations}
\begin{description}[
  labelindent=1em,
  labelwidth=1.2cm, 
  labelsep*=1em, 
  leftmargin = 3cm,
  itemindent= 0pt, 
  style = sameline,
  font=\normalfont\bfseries
]
    \item[AIL]
    Alamire Interactive Laboratory

    \item[CR]
    Crown Array

    \item[DH]
    Dummy Head

    \item[DR]
    Drum Track

    \item[FOA]
    First-Order Ambisonics

    \item[FPS]
    Frames per Second

    \item[MC1]
    Microphone Configuration 1

    \item[MC2]
    Microphone Configuration 2

    \item[MC3]
    Microphone Configuration 3

    \item[MDF]
    Medium-Density Fiberboard

    \item[MOV]
    Recordings During Motion

    \item[NCC]
    Normalized Cross-Correlation

    \item[PI]
    Piano Track

    \item[PS1]
    Perfect Sweeps up to \SI{1}{\kilo\hertz}

    \item[PS8]
    Perfect Sweeps up to \SI{8}{\kilo\hertz}

    \item[PSD]
    Power Spectral Density

    \item[RF]
    Reference Microphone
    
    \item[RIR]
    Room Impulse Response

    \item[SL]
    Left Loudspeaker

    \item[SP]
    Speech Track

    \item[SR]
    Right Loudspeaker

    \item[SRC]
    Source Signal

    \item[STAT]
    Stationary Recordings

    \item[UCA]
    Uniform Circular Array

    \item[ULA]
    Uniform Linear Array

    \item[WN]
    White Noise Track

\end{description}

\bibliography{references}


\begin{thebibliography}{51}
\ifx \bisbn   \undefined \def \bisbn  #1{ISBN #1}\fi
\ifx \binits  \undefined \def \binits#1{#1}\fi
\ifx \bauthor  \undefined \def \bauthor#1{#1}\fi
\ifx \batitle  \undefined \def \batitle#1{#1}\fi
\ifx \bjtitle  \undefined \def \bjtitle#1{#1}\fi
\ifx \bvolume  \undefined \def \bvolume#1{\textbf{#1}}\fi
\ifx \byear  \undefined \def \byear#1{#1}\fi
\ifx \bissue  \undefined \def \bissue#1{#1}\fi
\ifx \bfpage  \undefined \def \bfpage#1{#1}\fi
\ifx \blpage  \undefined \def \blpage #1{#1}\fi
\ifx \burl  \undefined \def \burl#1{\textsf{#1}}\fi
\ifx \doiurl  \undefined \def \doiurl#1{\url{https://doi.org/#1}}\fi
\ifx \betal  \undefined \def \betal{\textit{et al.}}\fi
\ifx \binstitute  \undefined \def \binstitute#1{#1}\fi
\ifx \binstitutionaled  \undefined \def \binstitutionaled#1{#1}\fi
\ifx \bctitle  \undefined \def \bctitle#1{#1}\fi
\ifx \beditor  \undefined \def \beditor#1{#1}\fi
\ifx \bpublisher  \undefined \def \bpublisher#1{#1}\fi
\ifx \bbtitle  \undefined \def \bbtitle#1{#1}\fi
\ifx \bedition  \undefined \def \bedition#1{#1}\fi
\ifx \bseriesno  \undefined \def \bseriesno#1{#1}\fi
\ifx \blocation  \undefined \def \blocation#1{#1}\fi
\ifx \bsertitle  \undefined \def \bsertitle#1{#1}\fi
\ifx \bsnm \undefined \def \bsnm#1{#1}\fi
\ifx \bsuffix \undefined \def \bsuffix#1{#1}\fi
\ifx \bparticle \undefined \def \bparticle#1{#1}\fi
\ifx \barticle \undefined \def \barticle#1{#1}\fi
\bibcommenthead
\ifx \bconfdate \undefined \def \bconfdate #1{#1}\fi
\ifx \botherref \undefined \def \botherref #1{#1}\fi
\ifx \url \undefined \def \url#1{\textsf{#1}}\fi
\ifx \bchapter \undefined \def \bchapter#1{#1}\fi
\ifx \bbook \undefined \def \bbook#1{#1}\fi
\ifx \bcomment \undefined \def \bcomment#1{#1}\fi
\ifx \oauthor \undefined \def \oauthor#1{#1}\fi
\ifx \citeauthoryear \undefined \def \citeauthoryear#1{#1}\fi
\ifx \endbibitem  \undefined \def \endbibitem {}\fi
\ifx \bconflocation  \undefined \def \bconflocation#1{#1}\fi
\ifx \arxivurl  \undefined \def \arxivurl#1{\textsf{#1}}\fi
\csname PreBibitemsHook\endcsname

\bibitem[\protect\citeauthoryear{MacWilliam et~al.}{2024}]{macwilliamStatespaceEstimationSpatially2024}
\begin{botherref}
\oauthor{\bsnm{MacWilliam}, \binits{K.}},
\oauthor{\bsnm{Dietzen}, \binits{T.}},
\oauthor{\bsnm{Ali}, \binits{R.}},
\oauthor{\bsnm{{van Waterschoot}}, \binits{T.}}:
State-space estimation of spatially dynamic room impulse responses using a room acoustic model-based prior.
Frontiers Signal Process.
\textbf{4}
(2024)
\doiurl{10.3389/frsip.2024.1426082}
\end{botherref}
\endbibitem

\bibitem[\protect\citeauthoryear{Diaz-Guerra et~al.}{2021}]{diaz-guerraRobustSoundSource2021}
\begin{barticle}
\bauthor{\bsnm{Diaz-Guerra}, \binits{D.}},
\bauthor{\bsnm{Miguel}, \binits{A.}},
\bauthor{\bsnm{Beltran}, \binits{J.R.}}:
\batitle{Robust {Sound} {Source} {Tracking} {Using} {SRP}-{PHAT} and {3D} {Convolutional} {Neural} {Networks}}.
\bjtitle{IEEE/ACM Trans. Audio, Speech, Language Process.}
\bvolume{29},
\bfpage{300}--\blpage{311}
(\byear{2021})
\doiurl{10.1109/TASLP.2020.3040031}
\end{barticle}
\endbibitem

\bibitem[\protect\citeauthoryear{Ali and Christian}{2025}]{aliSource-timeDominantModeling2025}
\begin{botherref}
\oauthor{\bsnm{Ali}, \binits{R.}},
\oauthor{\bsnm{Christian}, \binits{A.}}:
Source-time dominant modeling of the doppler shift for the auralization of moving sources.
Acta Acust.
\textbf{9}
(2025)
\doiurl{10.1051/aacus/2024073}
\end{botherref}
\endbibitem

\bibitem[\protect\citeauthoryear{Quan and Li}{2024}]{quanMultichannelLongTermStreaming2024}
\begin{barticle}
\bauthor{\bsnm{Quan}, \binits{C.}},
\bauthor{\bsnm{Li}, \binits{X.}}:
\batitle{Multichannel {Long}-{Term} {Streaming} {Neural} {Speech} {Enhancement} for {Static} and {Moving} {Speakers}}.
\bjtitle{arXiv:2403.07675}
(\byear{2024})
\doiurl{10.48550/arXiv.2403.07675}
\end{barticle}
\endbibitem

\bibitem[\protect\citeauthoryear{Nophut et~al.}{2024}]{nophut2024velocity-controlled}
\begin{barticle}
\bauthor{\bsnm{Nophut}, \binits{M.}},
\bauthor{\bsnm{Preihs}, \binits{S.}},
\bauthor{\bsnm{Peissig}, \binits{J.}}:
\batitle{Velocity-controlled {Kalman} filter for an improved echo cancellation with continuously moving microphones}.
\bjtitle{J. Audio Eng. Soc.}
\bvolume{72},
\bfpage{33}--\blpage{43}
(\byear{2024})
\doiurl{10.17743/jaes.2022.0116}
\end{barticle}
\endbibitem

\bibitem[\protect\citeauthoryear{Katzberg et~al.}{2018}]{katzbergCompressedSensingFramework2018}
\begin{barticle}
\bauthor{\bsnm{Katzberg}, \binits{F.}},
\bauthor{\bsnm{Mazur}, \binits{R.}},
\bauthor{\bsnm{Maass}, \binits{M.}},
\bauthor{\bsnm{Koch}, \binits{P.}},
\bauthor{\bsnm{Mertins}, \binits{A.}}:
\batitle{A {Compressed} {Sensing} {Framework} for {Dynamic} {Sound}-{Field} {Measurements}}.
\bjtitle{IEEE/ACM Trans. Audio, Speech, Language Process.}
\bvolume{26}(\bissue{11}),
\bfpage{1962}--\blpage{1975}
(\byear{2018})
\doiurl{10.1109/TASLP.2018.2851144}
\end{barticle}
\endbibitem

\bibitem[\protect\citeauthoryear{{van Waterschoot}}{2025}]{vanWaterschootDeepDataDriven2025}
\begin{bchapter}
\bauthor{\bsnm{{van Waterschoot}}, \binits{T.}}:
\bctitle{Deep, data-driven modeling of room acoustics: literature review and research perspectives}.
In: \bbtitle{Proc. 11th Convent. Europ. Acoust. Assoc. Forum Acusticum/EuroNoise 2025},
\bconflocation{Málaga, Spain},
pp. \bfpage{4065}--\blpage{4072}
(\byear{2025}).
\doiurl{10.61782/fa.2025.0264}
\end{bchapter}
\endbibitem

\bibitem[\protect\citeauthoryear{Zheng et~al.}{2024}]{zhengBATLearningReason2024}
\begin{bchapter}
\bauthor{\bsnm{Zheng}, \binits{Z.}},
\bauthor{\bsnm{Peng}, \binits{P.}},
\bauthor{\bsnm{Ma}, \binits{Z.}},
\bauthor{\bsnm{Chen}, \binits{X.}},
\bauthor{\bsnm{Choi}, \binits{E.}},
\bauthor{\bsnm{Harwath}, \binits{D.}}:
\bctitle{{BAT}: {Learning} to {Reason} about {Spatial} {Sounds} with {Large} {Language} {Models}}.
In: \bbtitle{Proc. 41st Int. Conf. Machine Learn. (ICML)},
\bconflocation{Vienna, Austria}
(\byear{2024}).
\doiurl{10.48550/arXiv.2402.01591}
\end{bchapter}
\endbibitem

\bibitem[\protect\citeauthoryear{He et~al.}{2024}]{heDeepNeuralRoom2024}
\begin{bchapter}
\bauthor{\bsnm{He}, \binits{Y.}},
\bauthor{\bsnm{Cherian}, \binits{A.}},
\bauthor{\bsnm{Wichern}, \binits{G.}},
\bauthor{\bsnm{Markham}, \binits{A.}}:
\bctitle{Deep {Neural} {Room} {Acoustics} {Primitive}}.
In: \bbtitle{Proc. 41st Int. Conf. Machine Learn. (ICML)},
\bconflocation{Vienna, Austria},
pp. \bfpage{17842}--\blpage{17857}
(\byear{2024}).
\burl{https://proceedings.mlr.press/v235/he24b.html}
\end{bchapter}
\endbibitem

\bibitem[\protect\citeauthoryear{Nielsen et~al.}{2014}]{nielsenSingleMultichannelAudio2014}
\begin{bchapter}
\bauthor{\bsnm{Nielsen}, \binits{J.K.}},
\bauthor{\bsnm{Jensen}, \binits{J.R.}},
\bauthor{\bsnm{Jensen}, \binits{S.H.}},
\bauthor{\bsnm{Christensen}, \binits{M.G.}}:
\bctitle{The single- and multichannel audio recordings database ({SMARD})}.
In: \bbtitle{Proc. 14th {Int.} {Workshop} {Acoust.} {Signal} {Enhancement} ({IWAENC})},
\bconflocation{Juan-les-Pins},
pp. \bfpage{40}--\blpage{44}
(\byear{2014}).
\doiurl{10.1109/IWAENC.2014.6953334}
\end{bchapter}
\endbibitem

\bibitem[\protect\citeauthoryear{Eaton et~al.}{2015}]{eatonACEChallenge20142015}
\begin{bchapter}
\bauthor{\bsnm{Eaton}, \binits{J.}},
\bauthor{\bsnm{Gaubitch}, \binits{N.D.}},
\bauthor{\bsnm{Moore}, \binits{A.H.}},
\bauthor{\bsnm{Naylor}, \binits{P.A.}}:
\bctitle{The {ACE} challenge 2014; {Corpus} description and performance evaluation}.
In: \bbtitle{Proc. 2015 {IEEE} {Workshop} {Appls.} {Signal} {Process.} {Audio} {Acoust.} ({WASPAA})},
\bconflocation{New Paltz, NY, USA},
pp. \bfpage{1}--\blpage{5}
(\byear{2015}).
\doiurl{10.1109/WASPAA.2015.7336912}
\end{bchapter}
\endbibitem

\bibitem[\protect\citeauthoryear{Woods et~al.}{2015}]{woodsRealworldRecordingDatabase2015}
\begin{bchapter}
\bauthor{\bsnm{Woods}, \binits{W.S.}},
\bauthor{\bsnm{Hadad}, \binits{E.}},
\bauthor{\bsnm{Merks}, \binits{I.}},
\bauthor{\bsnm{Xu}, \binits{B.}},
\bauthor{\bsnm{Gannot}, \binits{S.}},
\bauthor{\bsnm{Zhang}, \binits{T.}}:
\bctitle{A real-world recording database for ad hoc microphone arrays}.
In: \bbtitle{Proc. 2015 {IEEE} {Workshop} {Appls.} {Signal} {Process.} {Audio} {Acoust.} ({WASPAA})},
\bconflocation{New Paltz, NY, USA},
pp. \bfpage{1}--\blpage{5}
(\byear{2015}).
\doiurl{10.1109/WASPAA.2015.7336915}
\end{bchapter}
\endbibitem

\bibitem[\protect\citeauthoryear{Sheelvant et~al.}{2019}]{sheelvantRSL2019RealisticSpeech2019}
\begin{bchapter}
\bauthor{\bsnm{Sheelvant}, \binits{R.}},
\bauthor{\bsnm{Sharma}, \binits{B.}},
\bauthor{\bsnm{Madhavi}, \binits{M.}},
\bauthor{\bsnm{Das}, \binits{R.K.}},
\bauthor{\bsnm{Prasanna}, \binits{S.R.M.}},
\bauthor{\bsnm{Li}, \binits{H.}}:
\bctitle{{RSL2019}: {A} {Realistic} {Speech} {Localization} {Corpus}}.
In: \bbtitle{Proc. 22nd {Conf.} {Oriental} {COCOSDA} {Int.} {Committee} {Co}-ordination {Standardisation} {Speech} {Databases} {Assessment} {Techniques} ({O}-{COCOSDA})},
pp. \bfpage{1}--\blpage{6}.
\bpublisher{IEEE},
\blocation{Cebu, Philippines}
(\byear{2019}).
\doiurl{10.1109/O-COCOSDA46868.2019.9060842}
\end{bchapter}
\endbibitem

\bibitem[\protect\citeauthoryear{Dietzen et~al.}{2023}]{dietzenMYRiAD2023}
\begin{botherref}
\oauthor{\bsnm{Dietzen}, \binits{T.}},
\oauthor{\bsnm{Ali}, \binits{R.}},
\oauthor{\bsnm{Taseska}, \binits{M.}},
\oauthor{\bsnm{{van Waterschoot}}, \binits{T.}}:
{MYRiAD}: a multi-array room acoustic database.
EURASIP J. Audio, Speech Music Process.
\textbf{17}
(2023)
\doiurl{10.1186/s13636-023-00284-9}
\end{botherref}
\endbibitem

\bibitem[\protect\citeauthoryear{Stewart and Sandler}{2010}]{stewartDatabaseOmnidirectionalBformat2010}
\begin{bchapter}
\bauthor{\bsnm{Stewart}, \binits{R.}},
\bauthor{\bsnm{Sandler}, \binits{M.}}:
\bctitle{Database of omnidirectional and {B}-format room impulse responses}.
In: \bbtitle{Proc. 2010 {IEEE} {Int.} {Conf.} {Acoust.}, {Speech} {Signal} {Process.} (ICASSP)},
\bconflocation{Dallas, TX, USA},
pp. \bfpage{165}--\blpage{168}
(\byear{2010}).
\doiurl{10.1109/ICASSP.2010.5496083}
\end{bchapter}
\endbibitem

\bibitem[\protect\citeauthoryear{Hadad et~al.}{2014}]{hadadMultichannelAudioDatabase2014}
\begin{bchapter}
\bauthor{\bsnm{Hadad}, \binits{E.}},
\bauthor{\bsnm{Heese}, \binits{F.}},
\bauthor{\bsnm{Vary}, \binits{P.}},
\bauthor{\bsnm{Gannot}, \binits{S.}}:
\bctitle{Multichannel audio database in various acoustic environments}.
In: \bbtitle{Proc. 14th {Int.} {Workshop} {Acoust.} {Signal} {Enhancement} ({IWAENC})},
\bconflocation{Juan-les-Pins, France}
(\byear{2014}).
\doiurl{10.1109/IWAENC.2014.6954309}
\end{bchapter}
\endbibitem

\bibitem[\protect\citeauthoryear{Čmejla et~al.}{2021}]{cmejlaMIRaGeMultichannelDatabase2021}
\begin{bchapter}
\bauthor{\bsnm{Čmejla}, \binits{J.}},
\bauthor{\bsnm{Kounovský}, \binits{T.}},
\bauthor{\bsnm{Gannot}, \binits{S.}},
\bauthor{\bsnm{Koldovský}, \binits{Z.}},
\bauthor{\bsnm{Tandeitnik}, \binits{P.}}:
\bctitle{Mirage: Multichannel database of room impulse responses measured on high-resolution cube-shaped grid}.
In: \bbtitle{Proc. 28th European Signal Process. Conf. (EUSIPCO)},
pp. \bfpage{56}--\blpage{60}
(\byear{2021}).
\doiurl{10.23919/Eusipco47968.2020.9287646}
\end{bchapter}
\endbibitem

\bibitem[\protect\citeauthoryear{Venkatakrishnan et~al.}{2021}]{venkatakrishnanTampereUniversityRotated2021}
\begin{bchapter}
\bauthor{\bsnm{Venkatakrishnan}, \binits{A.V.}},
\bauthor{\bsnm{Pertila}, \binits{P.}},
\bauthor{\bsnm{Parviainen}, \binits{M.}}:
\bctitle{Tampere {University} {Rotated} {Circular} {Array} {Dataset}}.
In: \bbtitle{Proc. 29th {European} {Signal} {Process.} {Conf.} ({EUSIPCO})},
\bconflocation{Dublin, Ireland},
pp. \bfpage{201}--\blpage{205}
(\byear{2021}).
\doiurl{10.23919/EUSIPCO54536.2021.9616072}
\end{bchapter}
\endbibitem

\bibitem[\protect\citeauthoryear{Koyama et~al.}{2021}]{koyamaMESHRIRDatasetRoom2021}
\begin{bchapter}
\bauthor{\bsnm{Koyama}, \binits{S.}},
\bauthor{\bsnm{Nishida}, \binits{T.}},
\bauthor{\bsnm{Kimura}, \binits{K.}},
\bauthor{\bsnm{Abe}, \binits{T.}},
\bauthor{\bsnm{Ueno}, \binits{N.}},
\bauthor{\bsnm{Brunnstrom}, \binits{J.}}:
\bctitle{{MESHRIR}: {A} {Dataset} of {Room} {Impulse} {Responses} on {Meshed} {Grid} {Points} for {Evaluating} {Sound} {Field} {Analysis} and {Synthesis} {Methods}}.
In: \bbtitle{Proc. 2021 {IEEE} {Workshop} {Appls.} {Signal} {Process.} {Audio} {Acoust.} ({WASPAA})},
\bconflocation{New Paltz, NY, USA},
pp. \bfpage{1}--\blpage{5}
(\byear{2021}).
\doiurl{10.1109/WASPAA52581.2021.9632672}
\end{bchapter}
\endbibitem

\bibitem[\protect\citeauthoryear{McKenzie et~al.}{2021}]{mckenzie2021datasetspatialroomimpulse}
\begin{barticle}
\bauthor{\bsnm{McKenzie}, \binits{T.}},
\bauthor{\bsnm{McCormack}, \binits{L.}},
\bauthor{\bsnm{Hold}, \binits{C.}}:
\batitle{Dataset of spatial room impulse responses in a variable acoustics room for six degrees-of-freedom rendering and analysis}.
\bjtitle{arXiv:2111.11882}
(\byear{2021})
\doiurl{10.48550/arXiv.2111.11882}
\end{barticle}
\endbibitem

\bibitem[\protect\citeauthoryear{Zhao et~al.}{2022}]{zhaoRoomImpulseResponse2022}
\begin{barticle}
\bauthor{\bsnm{Zhao}, \binits{S.}},
\bauthor{\bsnm{Zhu}, \binits{Q.}},
\bauthor{\bsnm{Cheng}, \binits{E.}},
\bauthor{\bsnm{Burnett}, \binits{I.S.}}:
\batitle{A room impulse response database for multizone sound field reproduction ({L})}.
\bjtitle{J. Acoust. Soc. Amer.}
\bvolume{152}(\bissue{4}),
\bfpage{2505}--\blpage{2512}
(\byear{2022})
\doiurl{10.1121/10.0014958}
\end{barticle}
\endbibitem

\bibitem[\protect\citeauthoryear{Miotello et~al.}{2024}]{miotelloHOMULARIRRoomImpulse2024}
\begin{bchapter}
\bauthor{\bsnm{Miotello}, \binits{F.}},
\bauthor{\bsnm{Ostan}, \binits{P.}},
\bauthor{\bsnm{Pezzoli}, \binits{M.}},
\bauthor{\bsnm{Comanducci}, \binits{L.}},
\bauthor{\bsnm{Bernardini}, \binits{A.}},
\bauthor{\bsnm{Antonacci}, \binits{F.}},
\bauthor{\bsnm{Sarti}, \binits{A.}}:
\bctitle{{HOMULA-RIR}: A room impulse response dataset for teleconferencing and spatial audio applications acquired through higher-order microphones and uniform linear microphone arrays}.
In: \bbtitle{Proc. 2024 IEEE Int. Conf. Acoust. Speech Signal Process. Workshops (ICASSPW)},
pp. \bfpage{795}--\blpage{799}
(\byear{2024}).
\doiurl{10.1109/ICASSPW62465.2024.10626753}
\end{bchapter}
\endbibitem

\bibitem[\protect\citeauthoryear{Kujawski et~al.}{2024}]{kujawskiMIRACLEMicrophoneArray2024}
\begin{barticle}
\bauthor{\bsnm{Kujawski}, \binits{A.}},
\bauthor{\bsnm{Pelling}, \binits{A.J.R.}},
\bauthor{\bsnm{Sarradj}, \binits{E.}}:
\batitle{{MIRACLE}—a microphone array impulse response dataset for acoustic learning}.
\bjtitle{EURASIP J. Audio, Speech Music Process.}
\bvolume{2024}(\bissue{1}),
\bfpage{32}
(\byear{2024})
\doiurl{10.1186/s13636-024-00352-8}
\end{barticle}
\endbibitem

\bibitem[\protect\citeauthoryear{Treybig et~al.}{2024}]{treybigHighSpatialResolution2024}
\begin{bchapter}
\bauthor{\bsnm{Treybig}, \binits{L.}},
\bauthor{\bsnm{Klein}, \binits{F.}},
\bauthor{\bsnm{Stolz}, \binits{G.}},
\bauthor{\bsnm{Werner}, \binits{S.}},
\bauthor{\bsnm{Gari}, \binits{S.V.A.}}:
\bctitle{A {High} {Spatial} {Resolution} {Dataset} of {Spatial} {Room} {Impulse} {Responses} for {Different} {Acoustic} {Room} {Configurations}}.
In: \bbtitle{Proc. 50th {Jahrestagung} Für {Akustik} ({DAGA})},
\bconflocation{Hannover, Germany},
pp. \bfpage{248}--\blpage{250}
(\byear{2024})
\end{bchapter}
\endbibitem

\bibitem[\protect\citeauthoryear{He et~al.}{2018}]{heDeepNeuralNetworks2018a}
\begin{bchapter}
\bauthor{\bsnm{He}, \binits{W.}},
\bauthor{\bsnm{Motlicek}, \binits{P.}},
\bauthor{\bsnm{Odobez}, \binits{J.-M.}}:
\bctitle{Deep {Neural} {Networks} for {Multiple} {Speaker} {Detection} and {Localization}}.
In: \bbtitle{Proc. 2018 {IEEE} {Int.} {Conf.} {Robotics} {Automation} ({ICRA})},
pp. \bfpage{74}--\blpage{79}.
\bpublisher{IEEE},
\blocation{Brisbane, QLD}
(\byear{2018}).
\doiurl{10.1109/ICRA.2018.8461267}
\end{bchapter}
\endbibitem

\bibitem[\protect\citeauthoryear{Strauss et~al.}{2018}]{straussDREGONDatasetMethods2018}
\begin{bchapter}
\bauthor{\bsnm{Strauss}, \binits{M.}},
\bauthor{\bsnm{Mordel}, \binits{P.}},
\bauthor{\bsnm{Miguet}, \binits{V.}},
\bauthor{\bsnm{Deleforge}, \binits{A.}}:
\bctitle{{DREGON}: {Dataset} and {Methods} for {UAV}-{Embedded} {Sound} {Source} {Localization}}.
In: \bbtitle{Proc. 2018 {IEEE}/{RSJ} {Int.} {Conf.} {Intell.} {Robots} {Syst.} ({IROS})},
\bconflocation{Madrid},
pp. \bfpage{1}--\blpage{8}
(\byear{2018}).
\doiurl{10.1109/IROS.2018.8593581}
\end{bchapter}
\endbibitem

\bibitem[\protect\citeauthoryear{Evers et~al.}{2020}]{eversLOCATAChallengeAcoustic2020}
\begin{barticle}
\bauthor{\bsnm{Evers}, \binits{C.}},
\bauthor{\bsnm{Lollmann}, \binits{H.W.}},
\bauthor{\bsnm{Mellmann}, \binits{H.}},
\bauthor{\bsnm{Schmidt}, \binits{A.}},
\bauthor{\bsnm{Barfuss}, \binits{H.}},
\bauthor{\bsnm{Naylor}, \binits{P.A.}},
\bauthor{\bsnm{Kellermann}, \binits{W.}}:
\batitle{The {LOCATA} {Challenge}: {Acoustic} {Source} {Localization} and {Tracking}}.
\bjtitle{IEEE/ACM Trans. Audio, Speech Language Process.}
\bvolume{28},
\bfpage{1620}--\blpage{1643}
(\byear{2020})
\doiurl{10.1109/TASLP.2020.2990485}
\end{barticle}
\endbibitem

\bibitem[\protect\citeauthoryear{Politis et~al.}{2020}]{politisDatasetReverberantSpatial2020a}
\begin{barticle}
\bauthor{\bsnm{Politis}, \binits{A.}},
\bauthor{\bsnm{Adavanne}, \binits{S.}},
\bauthor{\bsnm{Virtanen}, \binits{T.}}:
\batitle{A {Dataset} of {Reverberant} {Spatial} {Sound} {Scenes} with {Moving} {Sources} for {Sound} {Event} {Localization} and {Detection}}.
\bjtitle{arXiv:2006.01919}
(\byear{2020})
\doiurl{10.48550/arXiv.2006.01919}
\end{barticle}
\endbibitem

\bibitem[\protect\citeauthoryear{Politis et~al.}{2022}]{politisSTARSS22DatasetSpatial2022}
\begin{barticle}
\bauthor{\bsnm{Politis}, \binits{A.}},
\bauthor{\bsnm{Shimada}, \binits{K.}},
\bauthor{\bsnm{Sudarsanam}, \binits{P.}},
\bauthor{\bsnm{Adavanne}, \binits{S.}},
\bauthor{\bsnm{Krause}, \binits{D.}},
\bauthor{\bsnm{Koyama}, \binits{Y.}},
\bauthor{\bsnm{Takahashi}, \binits{N.}},
\bauthor{\bsnm{Takahashi}, \binits{S.}},
\bauthor{\bsnm{Mitsufuji}, \binits{Y.}},
\bauthor{\bsnm{Virtanen}, \binits{T.}}:
\batitle{{STARSS22}: {A} dataset of spatial recordings of real scenes with spatiotemporal annotations of sound events}.
\bjtitle{arXiv:2206.01948}
(\byear{2022})
\doiurl{10.48550/arXiv.2206.01948}
\end{barticle}
\endbibitem

\bibitem[\protect\citeauthoryear{Brunnström et~al.}{2025}]{brunnstromExperimental2025}
\begin{bchapter}
\bauthor{\bsnm{Brunnström}, \binits{J.}},
\bauthor{\bsnm{Møller}, \binits{M.B.}},
\bauthor{\bsnm{Waterschoot}, \binits{T.}},
\bauthor{\bsnm{Moonen}, \binits{M.}},
\bauthor{\bsnm{Østergaard}, \binits{J.}}:
\bctitle{Experimental validation of sound field estimation methods using moving microphones}.
In: \bbtitle{Proc. 11th Convent. Europ. Acoust. Assoc. Forum Acusticum/EuroNoise 2025},
\bconflocation{Malaga, Spain},
pp. \bfpage{4111}--\blpage{4118}
(\byear{2025}).
\doiurl{10.61782/fa.2025.0379}
\end{bchapter}
\endbibitem

\bibitem[\protect\citeauthoryear{Lathoud et~al.}{2005}]{lathoudAV163AudioVisualCorpus2005}
\begin{bchapter}
\bauthor{\bsnm{Lathoud}, \binits{G.}},
\bauthor{\bsnm{Odobez}, \binits{J.-M.}},
\bauthor{\bsnm{Gatica-Perez}, \binits{D.}}:
\bctitle{{AV16}.3: {An} {Audio}-{Visual} {Corpus} for {Speaker} {Localization} and {Tracking}}.
In: \bbtitle{Machine {Learning} for {Multimodal} {Interaction}}
vol. \bseriesno{3361},
pp. \bfpage{182}--\blpage{195}.
\bpublisher{Springer},
\blocation{Berlin, Heidelberg}
(\byear{2005}).
\doiurl{10.1007/978-3-540-30568-2_16}
\end{bchapter}
\endbibitem

\bibitem[\protect\citeauthoryear{Deleforge and Horaud}{2012}]{deleforgeLatentlyConstrainedMixture2012}
\begin{bchapter}
\bauthor{\bsnm{Deleforge}, \binits{A.}},
\bauthor{\bsnm{Horaud}, \binits{R.}}:
\bctitle{A {Latently} {Constrained} {Mixture} {Model} for {Audio} {Source} {Separation} and {Localization}}.
In: \bbtitle{Latent {Variable} {Analysis} and {Signal} {Separation}}
vol. \bseriesno{7191},
pp. \bfpage{372}--\blpage{379}.
\bpublisher{Springer},
\blocation{Berlin, Heidelberg}
(\byear{2012}).
\doiurl{10.1007/978-3-642-28551-6_46}
\end{bchapter}
\endbibitem

\bibitem[\protect\citeauthoryear{Alameda-Pineda et~al.}{2013}]{alameda-pinedaRAVELAnnotatedCorpus2013}
\begin{barticle}
\bauthor{\bsnm{Alameda-Pineda}, \binits{X.}},
\bauthor{\bsnm{Sanchez-Riera}, \binits{J.}},
\bauthor{\bsnm{Wienke}, \binits{J.}},
\bauthor{\bsnm{Franc}, \binits{V.}},
\bauthor{\bsnm{Čech}, \binits{J.}},
\bauthor{\bsnm{Kulkarni}, \binits{K.}},
\bauthor{\bsnm{Deleforge}, \binits{A.}},
\bauthor{\bsnm{Horaud}, \binits{R.}}:
\batitle{{RAVEL}: an annotated corpus for training robots with audiovisual abilities}.
\bjtitle{J. Multimodal User Interfaces}
\bvolume{7}(\bissue{1-2}),
\bfpage{79}--\blpage{91}
(\byear{2013})
\doiurl{10.1007/s12193-012-0111-y}
\end{barticle}
\endbibitem

\bibitem[\protect\citeauthoryear{Allen and Berkley}{1979}]{allenImageMethodEfficiently1979}
\begin{barticle}
\bauthor{\bsnm{Allen}, \binits{J.B.}},
\bauthor{\bsnm{Berkley}, \binits{D.A.}}:
\batitle{Image method for efficiently simulating small-room acoustics}.
\bjtitle{J. Acoust. Soc. Amer.}
\bvolume{65}(\bissue{4}),
\bfpage{943}--\blpage{950}
(\byear{1979})
\doiurl{10.1121/1.382599}
\end{barticle}
\endbibitem

\bibitem[\protect\citeauthoryear{Vorländer}{2020}]{vorlanderAuralizationFundamentalsAcoustics2020}
\begin{bbook}
\bauthor{\bsnm{Vorländer}, \binits{M.}}:
\bbtitle{Auralization: {Fundamentals} of {Acoustics}, {Modelling}, {Simulation}, {Algorithms} and {Acoustic} {Virtual} {Reality}}.
\bsertitle{{RWTHedition}}.
\bpublisher{Springer},
\blocation{Cham}
(\byear{2020}).
\doiurl{10.1007/978-3-030-51202-6}
\end{bbook}
\endbibitem

\bibitem[\protect\citeauthoryear{{van Waterschoot}}{2022}]{lirias3940173}
\begin{bbook}
\bauthor{\bsnm{{van Waterschoot}}, \binits{T.}}:
\bbtitle{KU Leuven ESAT-STADIUS Audio Research Labs},
(\byear{2022}).
\bcomment{\url{https://lirias.kuleuven.be/3940173}}
\end{bbook}
\endbibitem

\bibitem[\protect\citeauthoryear{Damiano et~al.}{2025}]{trajectoRIR2024}
\begin{botherref}
\oauthor{\bsnm{Damiano}, \binits{S.}},
\oauthor{\bsnm{{MacWilliam}}, \binits{K.}},
\oauthor{\bsnm{Lorenzoni}, \binits{V.}},
\oauthor{\bsnm{Dietzen}, \binits{T.}},
\oauthor{\bsnm{{van Waterschoot}}, \binits{T.}}:
Data repository for the {trajectoRIR} database: room acoustic recordings along a trajectory of moving microphones
(2025).
\doiurl{10.5281/zenodo.15564430}
\end{botherref}
\endbibitem

\bibitem[\protect\citeauthoryear{Van~Oeteren and Eelen}{2023}]{alma9993576358501488}
\begin{botherref}
\oauthor{\bsnm{Van~Oeteren}, \binits{D.}},
\oauthor{\bsnm{Eelen}, \binits{K.}}:
Development of a mechatronic lab system for dynamic acoustic experiments.
Master's thesis,
KU Leuven, Faculty of Engineering Technology,
Leuven
(2023).
\url{https://kuleuven.limo.libis.be/discovery/fulldisplay?docid=alma9993576358501488&context=L&vid=32KUL_KUL:KULeuven&search_scope=All_Content&tab=all_content_tab&lang=en}
\end{botherref}
\endbibitem

\bibitem[\protect\citeauthoryear{Dokmanic et~al.}{2015}]{dokmanicEuclideanDistanceMatrices2015}
\begin{barticle}
\bauthor{\bsnm{Dokmanic}, \binits{I.}},
\bauthor{\bsnm{Parhizkar}, \binits{R.}},
\bauthor{\bsnm{Ranieri}, \binits{J.}},
\bauthor{\bsnm{Vetterli}, \binits{M.}}:
\batitle{Euclidean distance matrices: Essential theory, algorithms, and applications}.
\bjtitle{IEEE Signal Process. Mag.e}
\bvolume{32}(\bissue{6}),
\bfpage{12}--\blpage{30}
(\byear{2015})
\doiurl{10.1109/MSP.2015.2398954}
\end{barticle}
\endbibitem

\bibitem[\protect\citeauthoryear{Van~Oeteren and Eelen}{2023}]{vanOeterenThesisStadius2023}
\begin{botherref}
\oauthor{\bsnm{Van~Oeteren}, \binits{D.}},
\oauthor{\bsnm{Eelen}, \binits{K.}}:
thesisstadius.
GitHub repository: \url{https://github.com/DanteVanOeteren/thesisstadius/tree/main} (accessed 30/09/2024)
(2023)
\end{botherref}
\endbibitem

\bibitem[\protect\citeauthoryear{Holters et~al.}{2009}]{holtersImpulseResponseMeasurement2009}
\begin{bchapter}
\bauthor{\bsnm{Holters}, \binits{M.}},
\bauthor{\bsnm{Corbach}, \binits{T.}},
\bauthor{\bsnm{Zölzer}, \binits{U.}}:
\bctitle{Impulse {Response} {Measurement} {Techniques} and their {Applicability} in the {Real} {World}}.
In: \bbtitle{Proc. 12th {Int}. {Conf}. {Digital} {Audio} {Effects} ({DAFx09})},
\bconflocation{Como, Italy},
pp. \bfpage{108}--\blpage{112}
(\byear{2009})
\end{bchapter}
\endbibitem

\bibitem[\protect\citeauthoryear{Antweiler et~al.}{2012}]{Antweiler2012}
\begin{bchapter}
\bauthor{\bsnm{Antweiler}, \binits{C.}},
\bauthor{\bsnm{Telle}, \binits{A.}},
\bauthor{\bsnm{Vary}, \binits{P.}},
\bauthor{\bsnm{Enzner}, \binits{G.}}:
\bctitle{{Perfect-sweep NLMS for time-variant acoustic system identification}}.
In: \bbtitle{Proc. 2012 IEEE Int. Conf. Acoust., Speech, Signal Process. (ICASSP '12)},
pp. \bfpage{517}--\blpage{520}
(\byear{2012}).
\doiurl{10.1109/ICASSP.2012.6287930}
\end{bchapter}
\endbibitem

\bibitem[\protect\citeauthoryear{Simon}{2006}]{simon2006optimal}
\begin{botherref}
\oauthor{\bsnm{Simon}, \binits{D.}}:
Optimal state estimation: Kalman, H infinity, and nonlinear approaches.
John Wiley \& Sons
(2006)
\end{botherref}
\endbibitem

\bibitem[\protect\citeauthoryear{Postma and Katz}{2016}]{postma2016correction}
\begin{barticle}
\bauthor{\bsnm{Postma}, \binits{B.N.}},
\bauthor{\bsnm{Katz}, \binits{B.F.}}:
\batitle{Correction method for averaging slowly time-variant room impulse response measurements}.
\bjtitle{J. Acoust. Soc. Amer.}
\bvolume{140}(\bissue{1}),
\bfpage{38}--\blpage{43}
(\byear{2016})
\doiurl{10.1121/1.4955006}
\end{barticle}
\endbibitem

\bibitem[\protect\citeauthoryear{Prawda et~al.}{2023}]{prawda2023time}
\begin{bchapter}
\bauthor{\bsnm{Prawda}, \binits{K.}},
\bauthor{\bsnm{Schlecht}, \binits{S.J.}},
\bauthor{\bsnm{V{\"a}lim{\"a}ki}, \binits{V.}}:
\bctitle{Time variance in measured room impulse responses}.
In: \bbtitle{Proc.of Forum Acusticum 2023},
pp. \bfpage{1}--\blpage{8}
(\byear{2023}).
\doiurl{10.61782/fa.2023.0398}
\end{bchapter}
\endbibitem

\bibitem[\protect\citeauthoryear{Bhattacharjee et~al.}{2025}]{bhattacharjeeSoundSpeedPerturbation2025}
\begin{barticle}
\bauthor{\bsnm{Bhattacharjee}, \binits{S.S.}},
\bauthor{\bsnm{Jensen}, \binits{J.R.}},
\bauthor{\bsnm{Christensen}, \binits{M.G.}}:
\batitle{Sound speed perturbation robust audio: Impulse response correction and sound zone control}.
\bjtitle{IEEE Transactions on Audio, Speech and Language Processing}
\bvolume{33},
\bfpage{2008}--\blpage{2020}
(\byear{2025})
\doiurl{10.1109/TASLPRO.2025.3570949}
\end{barticle}
\endbibitem

\bibitem[\protect\citeauthoryear{Yamagishi et~al.}{2019}]{yamagishi_CSTRCorpus_2019}
\begin{botherref}
\oauthor{\bsnm{Yamagishi}, \binits{J.}},
\oauthor{\bsnm{Veaux}, \binits{C.}},
\oauthor{\bsnm{MacDonald}, \binits{K.}}:
CSTR VCTK Corpus: English Multi-speaker Corpus for CSTR Voice Cloning Toolkit (version 0.92).
University of Edinburgh. The Centre for Speech Technology Research (CSTR)
(2019).
\doiurl{10.7488/ds/2645}
\end{botherref}
\endbibitem

\bibitem[\protect\citeauthoryear{Anti-Everything}{2011}]{anti-everything_federation_2011}
\begin{botherref}
\oauthor{\bsnm{Anti-Everything}}:
Federation Day. Children of a Globalised World.
Musical Album, ISRC: TTA101100005,
(Boatshrimp Records, Port-of-Spain)
(2011)
\end{botherref}
\endbibitem

\bibitem[\protect\citeauthoryear{Kearney et~al.}{2009}]{kearney2009dynamic}
\begin{bchapter}
\bauthor{\bsnm{Kearney}, \binits{G.}},
\bauthor{\bsnm{Masterson}, \binits{C.}},
\bauthor{\bsnm{Adams}, \binits{S.}},
\bauthor{\bsnm{Boland}, \binits{F.}}:
\bctitle{Dynamic time warping for acoustic response interpolation: Possibilities and limitations}.
In: \bbtitle{Proc. 17th Europ. Signal Process. Conf. (EUSIPCO)},
pp. \bfpage{705}--\blpage{709}
(\byear{2009})
\end{bchapter}
\endbibitem

\bibitem[\protect\citeauthoryear{Enzner}{2010}]{enzner2010bayesian}
\begin{bchapter}
\bauthor{\bsnm{Enzner}, \binits{G.}}:
\bctitle{Bayesian inference model for applications of time-varying acoustic system identification}.
In: \bbtitle{Proc. 18th Europ. Signal Process. Conf. (EUSIPCO)},
pp. \bfpage{2126}--\blpage{2130}
(\byear{2010})
\end{bchapter}
\endbibitem

\bibitem[\protect\citeauthoryear{MacWilliam et~al.}{2025}]{macwilliam2025tracking}
\begin{bchapter}
\bauthor{\bsnm{MacWilliam}, \binits{K.}},
\bauthor{\bsnm{Dietzen}, \binits{T.}},
\bauthor{\bsnm{Waterschoot}, \binits{T.}}:
\bctitle{Tracking of spatially dynamic room impulse responses along locally linearized trajectories}.
In: \bbtitle{Proc. 11th Convent. Europ. Acoust. Assoc. Forum Acusticum/EuroNoise 2025},
\bconflocation{Malaga, Spain},
pp. \bfpage{4103}--\blpage{4110}
(\byear{2025}).
\doiurl{10.61782/fa.2025.0913}
\end{bchapter}
\endbibitem

\end{thebibliography}

\end{document}